\documentclass[prd,twocolumn,aps,superscriptaddress,nofootinbib,tightenlines]{revtex4}

\usepackage{amssymb}
\usepackage{braket}
\usepackage{bm}
\usepackage{pstricks,pst-node}

\newpsobject{showgrid}{psgrid}{subgriddiv=1,griddots=10,gridlabels=6pt}

{\count255=\time\divide\count255 by 60 \xdef\hourmin{\number\count255}
  \multiply\count255 by-60\advance\count255 by\time
  \xdef\hourmin{\hourmin:\ifnum\count255<10 0\fi\the\count255}}

\def\draftdate{{\rm timestamp:
\number\month/\number\day/\number\year\ \ \hourmin}}

\def\nn{ \nonumber \\ }

\def\vev#1{\left\langle #1 \right\rangle}

\def\Tr#1{\Braket{ #1 }} 
\def\tr#1{\braket{ #1 }}

\def\dd{d_D}
\def\dn{d_N}
\def\ng{{n_g}}
\def\mg{{n^\prime_g}}
\def\calU{\mathcal{U}}
\def\Uc{\mathsf{U}}
\def\U{\mathsf{U}}
\def\Um{\mathsf{U}^\prime}
\def\th{\vartheta}
\def\n{\mathcal{N}}

\begin{document}

\title{Algebraic Structure of Lepton and Quark Flavor Invariants and $CP$ Violation}

\author{Elizabeth E.~Jenkins}
\affiliation{Department of Physics, University of California at San Diego,
  La Jolla, CA 92093}

\author{Aneesh V.~Manohar}
\affiliation{Department of Physics, University of California at San Diego,
  La Jolla, CA 92093}

\date{\today\quad\hourmin}

\begin{abstract}
Lepton and quark flavor invariants are studied, both in the Standard Model with a dimension five Majorana neutrino mass operator, and in the seesaw model. The ring of invariants in the lepton sector is highly non-trivial, with non-linear relations among the basic invariants. The invariants are classified for the Standard Model with two and three generations, and for the seesaw model with two generations, and the Hilbert series is computed. The seesaw model with three generations proved computationally too difficult for a complete solution. We give an invariant definition of the $CP$-violating angle $\bar \vartheta$ in the electroweak sector.

\draftdate
\end{abstract}

\maketitle

\section{Introduction}

The observation of neutrino oscillations requires that the Standard Model~\cite{glashow,salam,weinbergsm} be modified to account for neutrino masses.  The leading theory of neutrino mass is the seesaw model~\cite{grs}, which contains additional fermions which are singlets under the $SU(3) \times SU(2) \times U(1)$ gauge group.   An attractive feature of the seesaw theory is that it naturally gives rise to leptogenesis~\cite{leptogenesis1} in $CP$-violating heavy neutrino decay.  The generated lepton asymmetry then produces a baryon asymmetry via Standard Model sphaleron processes.  Interestingly, the light neutrino masses favored by experiment are precisely in the range needed to produce a baryon asymmetry of the right magnitude~\cite{leptogenesis2,leptogenesis3}.  The baryon asymmetry is known to $\sim 10\%$ accuracy from the precision cosmic microwave background data of WMAP~\cite{wmap}.

At energies well below the mass scale $M$ of the heavy fermion singlets in the seesaw model, one constructs a low-energy effective theory obtained  by integrating out the heavy Majorana neutrino singlets.  The Lagrangian of the low-energy effective theory is given by the renormalizable Lagrangian of the Standard Model plus additional higher-dimensional terms obtained from integrating out the heavy neutrinos.  The leading term is a dimension-five operator~\cite{weinberg} which produces a Majorana mass term for the neutrinos of the Standard Model when the Higgs doublet acquires a vacuum expectation value $v$. This operator is the unique dimension-five operator which can be constructed from Standard Model fields.  Thus, it is natural for this dimension-five operator to be the first observed effect of new physics beyond the Standard Model.  The low-energy effective theory contains additional operators at dimension six~\cite{bgj1,bgj2}.  The leading effect of these operators is a flavor-nondiagonal correction to the weakly-interacting neutrino kinetic energy term after electroweak symmetry breakdown. This contribution results in a small ${\cal{O}}(v/M)$ nonunitary contribution to the lepton mixing matrix $U_{\text{PMNS}}$.  Unfortunately, for GUT-scale values of the seesaw scale $M$, this nonunitarity of  $U_{\text{PMNS}}$ is far too small to be observed experimentally.

 For the purposes of this paper, the Standard Model low-energy effective theory is the $SU(3) \times SU(2) \times U(1)$ gauge theory with only left-handed doublet neutrinos, plus an additional dimension-five gauge invariant operator which gives a Majorana mass to the neutrinos after spontaneous symmetry breaking, and the high-energy (renormalizable) theory is the seesaw model.

Flavor violation of quarks and leptons by Standard Model weak interactions is parameterized by unitary $3 \times 3$ matrices,  the CKM matrix in the quark sector and the PMNS matrix in the lepton sector. The fundamental parameters in the Standard Model are the quark and lepton Yukawa coupling matrices and the flavor matrix of the dimension-five Majorana mass operator. The fermion masses and mixing angles are derived quantities, obtained from the eigenvalues and eigenvectors of the flavor matrices in the low-energy theory. In the mass eigenstate basis, one still has the freedom to make phase rotations on the fermions fields, which leads to the redefinition of the CKM matrix
\begin{eqnarray}
V &\to& e^{-i \Phi_U} V e^{i \Phi_D} 
\label{rephase}
\end{eqnarray}
in the quark sector, where $\Phi_U=\text{diag}(\phi_u,\phi_c,\phi_t)$ and  $\Phi_D=\text{diag}(\phi_d,\phi_s,\phi_b)$. Physical quantities are basis independent, and must be invariant under the rephasing Eq.~(\ref{rephase}). CKM rephasing invariants have been studied extensively in the literature~\cite{jarlskog,greenberg,dunietz,jmrephasing}, the best-known example being the $CP$-odd Jarlskog invariant $J=\text{Im}\, V_{11}V_{22}V_{12}^* V_{21}^*$. Rephasing invariance also exists for the lepton mixing matrix.  In a previous paper~\cite{jmrephasing}, we extended the analysis of rephasing invariants to give a complete classification of these invariants for the Standard Model, and for the seesaw model.

The parameterization of the flavor structure in terms of masses and mixing angles is convenient for computing decay rates and scattering amplitudes. However, if one wants to understand the origin of flavor structure, the more fundamental quantities are the flavor matrices in the Lagrangian from which the masses and mixing angles are derived by diagonalization. A well-known difficulty is that the flavor matrices are basis-dependent, since one can make unitary transformations on the quark and lepton fields in the Lagrangian. 
For example, the Yukawa matrix for charge $2/3$ quarks transforms as 
\begin{eqnarray}\label{trans1}
Y_U &\to& {\U_{U^c}}^{T} \ Y_U \  \U_Q
\end{eqnarray}
where $\U_Q$ and $\U_{U^c}$ are unitary transformations on the quark doublet and singlet fields. One cannot directly compare a mass-matrix prediction with experiment, since the mass matrices are basis-dependent. Observable quantities must be independent of this change of basis, i.e.\ invariant under Eq.~(\ref{trans1}), and such quantities are sometimes referred to as weak basis invariants~\cite{Branco1,Branco2}. One can check the predictions of a flavor model by comparing invariant quantities with their corresponding experimental values.

Classifying invariants also is important in analyzing theories which explain flavor by a dynamical mechanism. The idea can be illustrated by a simple example --- consider a low-energy theory which has a $3\times 3$ hermitian traceless flavor matrix $X$ which transforms as an $SU(3)$ adjoint, $X \to  U X U^\dagger$. Imagine that $X$ is a dynamical variable in some high-energy theory, and that the low-energy value of $X$ is given by minimizing an effective potential $V(X)$ generated by the high-energy theory. It is well-known (see Sec.~\ref{sec:quarkinvariants}) that the only independent invariants are $I_2=\tr{X^2}$ and $I_3=\tr{X^3}$, where $\tr{*}$ denotes a matrix trace, so the potential can be written as $V(I_2,I_3)$ and minimizing it leads to the equation
\begin{eqnarray}\label{minv}
0 &=& \frac{\partial V}{\partial I_2}X^a + \frac34\frac{\partial V}{\partial I_3}
d_{abc} X^b X^c\,,
\end{eqnarray}
where $X=X^a T^a$. Eq.~(\ref{minv}) implies that $X^a =k\, d_{abc}X^b X^c$ where the constant of proportionality is $k=-(3/4) (\partial V/\partial I_3)/(\partial V/\partial I_2)$ evaluated at the minimum. The solutions of this equation are either (i) the trivial solution $X=0$, or (ii) $X$ can be brought to the diagonal form
\begin{eqnarray}
X = -\frac{1}{2k}
\left(
\begin{array}{ccc}
 1 & 0  & 0  \\
 0 &  1 & 0  \\
 0 & 0  & -2  
\end{array}
\right) = - {\sqrt{3} \over k} T^8
\end{eqnarray}
with symmetry breaking in the $T^8$ direction. Thus the $SU(3)$ symmetry is either (i) unbroken or (ii) broken to the $SU(2)\times U(1)$ subgroup. Symmetry breaking to $U(1)^3$ is not allowed. Examples of this type were studied in the early literature on unified theories~\cite{radicati,michel} in the context of understanding gauge symmetry breaking patterns by minimizing Higgs potentials. A recent example
from flavor physics needing the classification of invariants can be found in Ref.~\cite{feldmann}.

There is an extensive literature on quark and lepton invariants~(see, e.g. \cite{Branco1,Branco2,kusenko1,kusenko2,susy,kitano}). The main emphasis in previous work has been the study of $CP$ violation. $CP$-violating invariants analogous to the Jarlskog invariant were written down.  The vanishing of the $CP$-violating invariants was sufficient to guarantee the vanishing of $CP$ violation in the CKM and PMNS mixing matrices. 

In this paper, we take a different approach, studying all the invariants, and treating the problem using the methods of invariant theory~\cite{goodmanwallach,procesi,derksen}, which considers the ring of polynomials that are invariant under the action of a group. Polynomial invariants also are the relevant objects for physics applications, since an effective Lagrangian is written as a polynomial in the basic variables which describe the theory.\footnote{For example, the chiral Lagrangian is a polynomial in the quark mass matrix $M$.} A basic result of invariant theory is that the ring of invariants has a finite number of generators. There can be non-trivial relations among the invariants, known as syzygies~\cite{syzygy}, so that the invariant ring need not be a free ring. The number of invariants of a given degree is encoded in the Hilbert series. The complete classification of the invariant ring is, in general, a very difficult computational problem.

In this paper, we study the invariants of the Standard Model low-energy theory and the seesaw theory in both the quark and lepton sectors. In the quark sector, the complete structure of the invariant ring is given, and the relation between the polynomial invariants and rephasing invariants also is given. The structure of the invariant ring in the lepton sector is considerably more involved than in the quark sector. The classification of lepton  invariants is given for the low-energy effective Standard Model theory for two and three generations. For the high-energy seesaw theory, the classification is given for two generations. For three generations, we have been unable to completely classify all the relations or to determine the Hilbert series because the problem is computationally too difficult. The simpler invariants (i.e.\ of small degree) are given for this case.

The paper is organized as follows.  Section~\ref{sec:theories} defines the high-energy seesaw theory and its low-energy effective theory.  The flavor-symmetry breaking matrices and $\th$-angles of each theory are given, together with their transformation properties under flavor symmetry and $CP$.  Section~\ref{sec:parameters} defines the mass and mixing matrices of the high-energy and low-energy theories.  The high-energy theory contains three mixing matrices, the quark CKM mixing matrix $V_{\text{CKM}}$, its analogous lepton mixing matrix $V$ and a mixing matrix for the heavy neutrino singlets $W$.  The low-energy theory contains two mixing matrices, the quark CKM  mixing matrix $V_{\text{CKM}}$ and the lepton PMNS matrix $U_{\text{PMNS}}$.    Section~\ref{sec:parameters} explains the counting of mixing angles and phases for the mixing matrices for arbitrary numbers of Standard Model fermions and neutrino singlets in both the high- and low-energy theories.  Finally, rephasing invariance of the mixing matrices is discussed.  Section~\ref{sec:invariants} provides a brief introduction to the mathematics of invariant theory that we need for our analysis.  Several model theories are considered to elucidate the mathematical results.  The next two sections consider the classification of flavor invariants for the high-energy seesaw theory and its low-energy effective theory.  Section~\ref{sec:quarkinvariants} reviews the classification of the quark mass matrix invariants, which are identical to the quark invariants of the Standard Model for both theories.  Sections~\ref{sec:linv2gen} and~\ref{sec:linv3gen} consider the classification of lepton mass matrix invariants for two and three generations of fermions, respectively, in both the low-energy effective theory and the seesaw theory.  The complete classification is given for the low-energy effective theory for two and three generations.  The lepton invariant analysis of the full seesaw theory is significantly more complex.  The complete classification is given for two generations of fermions, and partial results for three generations are given.  

\section{Flavor Symmetries}\label{sec:theories}

We consider the $SU(3) \times SU(2) \times U(1)$ gauge theory with $\ng$ generations of Standard Model fermions and $\mg$ generations of gauge singlet fermions (neutrino singlets). The fermion multiplets are $Q_i=(\mathbf{3},\mathbf{2})_{1/6}$, $U^c_i=(\mathbf{\bar 3},\mathbf{1})_{-2/3}$, $D^c_i=(\mathbf{\bar 3},\mathbf{1})_{1/3}$, $L_i=(\mathbf{1},\mathbf{2})_{-1/2}$ and $E^c_i=(\mathbf{1},\mathbf{1})_{1}$, $i = 1, \ldots, \ng$, and $N^c_I=(\mathbf{1},\mathbf{1})_0$, $I= 1, \ldots, \mg$.  All fermion multiplets are left-handed Weyl fields.  The fermion multiplets with $\mg=\ng$ have a natural embedding in the ${\bf 16}$ spinor representation of $SO(10)$, so the usual choice is $\mg=\ng$. Theories with $\mg \neq \ng$ also are possible, however. Experimentally, we know that $\ng=3$, but there is no experimental limit on $\mg$. Big-bang nucleosynthesis constrains the number of neutrino flavors to be less than four; however, this only constrains neutrinos which are light enough to be present at temperatures of order an MeV.

The flavor symmetry of the fermion sector of the high-energy theory is $SU(\ng)^5 \times U(\mg) \times U(1)^2$, since there is a separate $SU(\ng)$ flavor symmetry for each of the five multiplets $Q$, $U^c$, $D^c$, $L$ and $E^c$, a $U(\mg)$ flavor symmetry for the singlets $N^c$, and two additional non-anomalous $U(1)$ flavor symmetries. Out of the six possible $U(1)$ symmetries, only three linear combinations are non-anomalous under $SU(3)\times SU(2) \times U(1)$: $N^c$ number which is included in $U(\mg)$, $(B-L)$, and $(E^c+D^c-U^c)$ number.  The three additional anomalous $U(1)$ groups can be treated as symmetries if the three $\th$-angles\footnote{The $\th$ angles multiplying $F \tilde F$ terms are not to be confused with angles $\theta$ of the quark and lepton mixing matrices. There are no instantons in the $U(1)$ sector, but the $\vartheta$ angle can have physical consquences in the presence of topological defects.} $\th_{3,2,1}$ of the $SU(3)$, $SU(2)$ and $U(1)$ gauge groups transform under arbitrary chiral phase transformations $\psi \to e^{i \alpha_\psi } \psi$ on the fields $\psi = Q,$ $U^c$, $D^c$, $L$ and $E^c$ as\begin{eqnarray}\label{thetatrans}
\th_3 &\to& \th_3 - \ng \left( 2\alpha_Q +\alpha_{U^c}+\alpha_{D^c} \right), \nn
\th_2 &\to& \th_2 - \ng \left( 3 \alpha_Q+\alpha_L \right), \\
\th_1 &\to& \th_1 - \ng \left( \frac16 \alpha_Q+\frac43 \alpha_{U^c}+\frac13 \alpha_{D^c} +\frac 12 \alpha_L+\alpha_{E^c} \right).\nonumber
\end{eqnarray}
Eq.~(\ref{thetatrans}) does not depend on $\mg$ or $\alpha_{N^c}$, since $N^c$ are gauge singlets. With the transformation Eq.~(\ref{thetatrans}), the chiral flavor symmetry becomes $U(\ng)^5 \times U(\mg)$, with a separate flavor factor for each of the six fermion multiplets.

The $U(\ng)^5 \times U(\mg)$ flavor symmetry of the fermion and gauge kinetic energy terms is explicitly broken by gauge-invariant renormalizable terms --- Yukawa couplings between fermion multiplets and the Higgs doublet and Majorana mass terms of the fermion singlets.  The flavor symmetry-breaking Lagrangian is given by
\begin{eqnarray}\label{high}
{\cal L} &=& - U^c_i  \left(Y_U\right)_{ij} Q_j H- D^c_i \left( Y_D \right)_{ij} Q_j H^\dagger \nn
&&-  E^c_i \left( Y_E \right)_{ij} L_j H^\dagger  
-  N^c_I \left( Y_\nu \right)_{Ij} L_j H \nn
&&-\frac 1 2 N^c_I M_{IJ} N^c_J +  \text{h.c.},
\end{eqnarray}
where $H=(1,2)_{1/2}$ is the Higgs doublet, and gauge and Lorentz indices have been suppressed. The Yukawa couplings $Y_{U,D,E}$ are $\ng \times \ng$ matrices, whereas the neutrino Yukawa coupling $Y_\nu$ is an $\mg \times \ng$ matrix.  The singlet neutrino Majorana mass matrix $M$ is a symmetric $\mg \times \mg$ matrix. In the Standard Model without neutrino singlets, renormalizable terms proportional to $Y_\nu$ and $M$ are absent.

Under the chiral flavor symmetry transformations $\psi \to \calU_{\psi}\ \psi$, where $\calU_{\psi}$ are unitary matrices in flavor space for the fermion fields $\psi= Q$, $U^c$, $D^c$, $L$, $E^c$ and $N^c$, the Yukawa coupling matrices, the Majorana mass matrix and the $\th$ angles transform as
\begin{eqnarray}\label{trans}
Y_U &\to& {\calU_{U^c}}^{T} \ Y_U \  \calU_Q , \nn
Y_D &\to& {\calU_{D^c}}^T \ Y_D \ \calU_Q , \nn
Y_E &\to& {\calU_{E^c}}^T \ Y_E \ \calU_L , \nn
Y_\nu &\to& {\calU_{N^c}}^T \ Y_\nu \ \calU_L , \nn
M &\to& {\calU_{N^c}}^T \  M \ \calU_{N^c} , \nn
\th_3 &\to& \th_3 - 2 \arg \det \calU_Q - \arg \det \calU_{U^c}-\arg \det \calU_{D^c} , \nn
\th_2 &\to& \th_2 - 3 \arg \det \calU_Q - \arg \det \calU_{L} , \nn
\th_1 &\to& \th_1 - \frac16 \arg \det \calU_Q - \frac43 \arg \det \calU_{U^c}-\frac13\arg \det \calU_{D^c}\nn
&&-\frac12\arg \det \calU_{L}-\arg \det \calU_{E^c} .
\end{eqnarray}
Under $CP$, each matrix is transformed to its complex conjugate, and each $\th$ angle changes sign, 
\begin{eqnarray}\label{cp}
Y_{U,D,E,\nu} &\to& Y_{U,D,E,\nu}^* \ ,\nn
M &\to& M^* \ ,\nn
\th_{1,2,3} &\to& - \th_{1,2,3} \ . 
\end{eqnarray}

Under the chiral flavor symmetry transformation, the $\th$ angles are shifted by Eq.~(\ref{trans}). The invariant angle $\bar \th_{\text{QCD}}$ is defined by
\begin{eqnarray}
\bar \th_{\text{QCD}} &=& \th_3 +  \arg \det Y_U+\arg \det Y_D\,. 
\label{thqcd}
\end{eqnarray} 
The analogous angles $\bar \th_{1,2}$ can not be separately defined, but one can define an invariant $\th$-parameter in the electroweak sector
\begin{eqnarray}
\bar \th_{\text{EW}} &=& \th_2 + 2 \th_1 + \frac83 \arg \det Y_U+\frac23\arg \det Y_D\nn
&&+2 \arg \det Y_E\,. 
\label{thew}
\end{eqnarray} 
After electroweak symmetry breaking, the QED $\vartheta$-angle is $2\bar \th_{\text{QED}}=\bar \th_{\text{EW}}$. The factor of two arises because the generators for a non-abelian gauge theory are normalized to $\text{Tr}\, T^a T^b=\delta^{ab}/2$.

In the absence of electroweak symmetry breaking, there are $\mg$ massive Majorana neutrino singlets with masses of $\mathcal{O}(M)$, the heavy Majorana neutrino mass scale, and all other fermions are strictly massless.  It is natural that $M$ be of order the GUT scale, the scale at which the GUT gauge symmetry breaks to the Standard Model gauge group, under which the $N^c$ fields are uncharged.  When the Higgs field gets a vacuum expectation value $v/\sqrt{2}$, the Yukawa matrices generate Dirac mass matrices for the quarks and leptons, 
\begin{eqnarray}
m_{U,D,E,\nu} &=& Y_{U,D,E,\nu} \ {v \over \sqrt{2}}\ .
\end{eqnarray}
with the same flavor transformation properties as the Yukawa couplings. The Dirac and Majorana mass matrices of the $(\ng + \mg)$ left-handed neutrino fields combine to form a neutrino mass term 
\begin{eqnarray}\label{fullmaj}
- \frac 1 2 \n_{\mathcal{I}} \left({M_\n}\right)_{\mathcal{I} \mathcal{J}} \n_{\mathcal{J}},\qquad 1 \le \mathcal{I},\mathcal{J} \le \ng+\mg
\end{eqnarray} 
where the $(\ng + \mg) \times (\ng + \mg)$ neutrino mass matrix $M_\n$ is equal to the symmetric matrix
\begin{eqnarray}\label{fullmajmass}
M_\n &\equiv& \left( \begin{array}{cc} 
0 &  {m_\nu}^T \\
{m_\nu} & M \end{array}\right) \ .
\end{eqnarray}
The $(\ng + \mg)$ neutrino fields $\n_{\mathcal{I}}$ are $(\nu_i,  N^c_I)$. The $(\ng + \mg)$ mass eigenstates of Eq.~(\ref{fullmajmass}) give the Majorana mass-eigenstate neutrino fields, which are linear combinations of $\nu_i$ and $N^c_I$. The heavy neutrinos with masses $\mathcal{O}(M)$ are predominantly $N^c$ with an $\mathcal{O}(v/M)$ admixture of $\nu$, and the light neutrinos with masses $\mathcal{O}(v^2/M)$ are predominantly $\nu$ with an $\mathcal{O}(v/M)$ admixture of $N^c$.

A low-energy effective field theory can be obtained from the seesaw theory by integrating out the $\mg$ heavy Majorana neutrino mass eigenstates.  In this low-energy theory, the leading flavor symmetry-breaking Lagrangian is given by
\begin{eqnarray}\label{low}
{\cal L}^{\text{EFT}} &=&  -U^c_i  \left(Y_U\right)_{ij} Q_j H- D^c_i \left( Y_D \right)_{ij} Q_j H^\dagger \nn
&&-  E^c_i \left( Y_E \right)_{ij} L_j H^\dagger \nn
&&+\frac 1 2 {(L_i H)} \left(C_5\right)_{ij} (L_j H) + \text{h.c.},
\end{eqnarray}
where the coefficient of the dimension-five operator~\cite{weinberg} is given by
\begin{eqnarray}\label{c5}
C_5 = Y_\nu^T \, {M}^{-1}\, {Y_\nu}
\end{eqnarray}
to lowest order in the $1/M$ expansion.  When the electroweak gauge symmetry breaks, the dimension-five operator yields an effective $\ng \times \ng$ Majorana mass matrix 
\begin{eqnarray}
m_5 = - C_5 v^2/2
\end{eqnarray}
for the (primarily) weak doublet neutrinos.  Under the flavor symmetries and $CP$, the flavor matrices $Y_{U,D,E}$ and $\th$ angles $\th_{1,2,3}$ of the low-energy effective theory transform under chiral flavor symmetry and $CP$ as in Eq.~(\ref{trans}) and Eq.~(\ref{cp}), respectively, whereas $C_5$ transforms as
\begin{eqnarray}\label{c5trans}
C_5 &\to& {\calU_L}^T \ C_5 \ \calU_L , \nn
C_5 &\to& C_5^* \ ,
\end{eqnarray}  
respectively.

We will analyze the flavor structure of both the seesaw theory and its low-energy effective theory. The analysis depends only on the flavor transformation properties of the Yukawa coupling and Majorana mass matrices (i.e.\ the fermion mass matrices).  Thus, it applies to any theory which has Dirac and Majorana mass matrices with the same transformation properties as given here, regardless of whether the Dirac mass terms are  proportional to Yukawa couplings in the theory, or are generated by some mechanism from more fundamental parameters of the theory.

\section{Masses, Mixing Angles and Phases}\label{sec:parameters}

In this section, we define the mass and mixing parameters of the high-energy seesaw theory and its low-energy effective theory. Most of the section is a review of well-known results, and serves to define the parameters and notation which are needed later. The mass matrices of the high and low energy theories in the weak eigenstate basis are transformed  to the mass eigenstate basis by flavor rotations to obtain the fermion masses and mixing matrices. The counting of mixing angles and phases for the case $\mg = \ng$ follows the analysis of Ref.~\cite{jmrephasing}.  The counting of physical parameters is given here for the cases $\mg > \ng$ and $\mg < \ng$, for completeness.

Any complex matrix $M$ can be written in the form $M = \Uc\ \Lambda\ \U^\prime$ where $\Uc$ and $\U^\prime$ are unitary matrices, and $\Lambda$ is a diagonal matrix with real, non-negative entries.  If $M$ is also a symmetric matrix, then it can be written in the form  $M = M^T = {\U}^T\ \Lambda\ \U$, where $\U$ is a unitary matrix.

\subsection{High-Energy Theory}

The flavor matrices of the high-energy seesaw theory are written in Eq.~(\ref{high}) in the weak eigenstate basis.  These flavor matrices are related to the mass eigenstate basis by
\begin{eqnarray}\label{transhigh}
Y_U &=& \Uc_{U^c} \ \Lambda_U \  \U_U , \nn
Y_D &=& \Uc_{D^c} \ \Lambda_D \ \U_D , \nn
Y_E &=& \Uc_{E^c} \ \Lambda_E \ \U_E , \nn
Y_\nu &=& \Uc_{N^c}\ \Lambda_\nu \ \U_\nu , \nn
M &=& {\Um_{N^c}}^T \  \Lambda_{N} \ {\Um_{N^c}} ,
\end{eqnarray}
where $\Lambda_{U, D, E}$, $\Lambda_\nu$ and $\Lambda_N$ are $\ng \times \ng$, $\mg \times \ng$  and $\mg \times \mg$ diagonal matrices respectively, with real, non-negative entries;  $\Uc_{U^c,D^c,E^c}$ and $\U_{U,D,E,\nu}$ are $\ng \times \ng$ unitary matrices, and $\Uc_{N^c}$ and $\Um_{N^c}$ are $\mg \times \mg$ unitary matrices, which transform the mass eigenstate basis to the weak eigenstate basis.  Performing the chiral flavor transformation Eq.~(\ref{trans}) with ${\calU_{U^c}}^T={\Uc_U}^{-1}$, ${\calU_{D^c}}^T={\Uc_D}^{-1}$, ${\calU_{E^c}}^T= {\Uc_E}^{-1}$, $\calU_Q={\U_U}^{-1}$, $\calU_L={\U_E}^{-1}$, and $\calU_{N^c} = {\Um_{N^c}}^{-1}$ brings the flavor matrices to the form 
\begin{eqnarray}\label{diaghigh}
Y_U &=&  \Lambda_U , \nn
Y_D &=& \Lambda_D \ V_{\text{CKM}}^{-1} ,  \nn
Y_E &=& \Lambda_E , \nn
Y_\nu &=& W^{-1} \ \Lambda_\nu \ V , \nn
M  &=& \Lambda_N ,
\end{eqnarray}
where $V_{\text{CKM}} \equiv \U_U {\U_D}^{-1}$, $V \equiv \U_\nu {\U_E}^{-1}$ and $W \equiv {\Uc_{N^c}}^{-1} \left({\Um_{N^c}}\right)^T $ are the three unitary matrices which describe flavor mixing in the seesaw theory. $V_{\text{CKM}}$ is the Cabibbo-Kobayashi-Maskawa mixing matrix in the quark sector. As is well-known, this $\ng \times \ng$ matrix corresponds to the mismatch between the unitary field redefinitions on $U$ and $D$ in the quark doublets $Q$ required to diagonalize $Y_U$ and $Y_D$.  $V$ is the analogue of the CKM matrix in the lepton sector; it is the $\ng \times \ng$ matrix corresponding to the mismatch between the unitary field redefinitions on $\nu$ and $E$ in the lepton doublets $L$ required to diagonalize $Y_\nu$ and $Y_E$.  $W$ is an $\mg \times \mg$ mixing matrix in the lepton sector corresponding to the mismatch between the unitary field redefinitions on $N^c$ required to diagonalize $M$ and $Y_\nu$.

To proceed further, it is necessary to consider the three cases $\mg = \ng$, $\mg < \ng$ and $\mg > \ng$ individually. We first specialize to the case $\mg = \ng$ considered previously in Ref.~\cite{jmrephasing} and review the analysis given there. The analysis is then generalized to the cases $\mg \neq \ng$. The quark sector only depends on the number of quark generations $\ng$, but the lepton sector analysis depends on whether $\mg = \ng$, $\mg < \ng$ or $\mg > \ng$.  

\subsubsection{$\mg = \ng$}

The real diagonal matrices $\Lambda_{U,D,E,\nu,N}$ are invariant under the rephasings,
\begin{eqnarray}
\Lambda_{\psi} &\to& e^{-i \Phi_{\psi}} \ \Lambda_{\psi} \ e^{i \Phi_{\psi}} , \ \psi= U,D,E,\nn
\Lambda_\nu &\to& e^{-i \Phi_\nu} \ \Lambda_\nu \ e^{i \Phi_\nu},\nn
\Lambda_N &\to& \eta_N \ \Lambda_N \ \eta_N ,
\end{eqnarray}
where $\Phi_{U,D,E,\nu}$ are real diagonal matrices, and $\eta_N$ is a diagonal matrix with allowed eigenvalues $\pm 1$.  Only $\pm 1$ rephasings are allowed for the Majorana fields $N^c$.  Under these rephasings, the mixing matrices $V_{\text{CKM}}$, $V$ and $W$ transform as
\begin{eqnarray}\label{rephasinghigh}
V_{\text{CKM}} &\to& e^{-i \Phi_U}\ V_{\text{CKM}} \ e^{i \Phi_D},\nn
V &\to& e^{-i \Phi_\nu}\ V \ e^{i \Phi_E},\nn
W &\to& e^{-i \Phi_\nu}\ W\ \eta_N \,.
\end{eqnarray}

\medskip

\noindent \emph{Quark Sector:} The parameter counting in the quark sector is well-known, and is summarized here for completeness. The matrices $\Lambda_U$ and $\Lambda_D$ each contain $\ng$ eigenvalues, which correspond to the $U$-quark and $D$-quark masses, respectively, and are $CP$ even. The quark mixing matrix $V_{\text{CKM}}$ is an $\ng \times \ng$ unitary matrix with $\ng^2$ parameters. It is conventional to divide these parameters into angles and phases --- angles are even under $CP$, whereas phases are odd under $CP$. If the $V_{\text{CKM}}$ matrix is $CP$ invariant, it is an $\ng \times \ng$ real orthogonal matrix with $\ng(\ng-1)/2$ parameters. The unitary matrix $V_{\text{CKM}}$ has $\ng(\ng-1)/2$ angles and $\ng(\ng+1)/2$ phases, and can be parametrized by
\begin{eqnarray}
e^{i \chi}\ e^{i \Phi} \ \mathcal{V}(\theta_i, \delta_i)\ e^{i \Psi}\ ,
\end{eqnarray}
where $\chi$ is an overall phase, $\Phi= {\text{diag}}(0, \phi_2, \cdots, \phi_\ng)$, and $\Psi = {\text{diag}}(0, \psi_2, \cdots, \psi_\ng)$. The phase redefinitions $\Phi_U$ and $\Phi_D$ of $V_{\text{CKM}}$ in Eq.~(\ref{rephasinghigh}) can be chosen to remove the $2\ng-1$ phases $\chi$, $\phi_i$, $\psi_i$, $i=2, \cdots, \ng$.\footnote{There are $\ng$ phases each in $\Phi_U$ and $\Phi_D$, but the transformation $\Phi_U = \Phi_D \propto \openone$ leaves $V_{\text{CKM}}$ invariant.} Thus, $V_{\text{CKM}}$ has $\ng(\ng+1)/2 - (2\ng-1)=(\ng-1)(\ng-2)/2$ net phases. This counting of parameters is summarized in Table~I.

\begin{table}\label{tab:quark}
\begin{eqnarray*}
\begin{array}{c|ccc}
\text{Matrices}& \text{Masses} & \text{Angles} & \text{Phases} \\
\hline
\Lambda_U &  \ng & 0 & 0  \\
\Lambda_D &  \ng & 0 & 0  \\
V_{\text{CKM}} & 0 & \frac 1 2 \ng(\ng-1) & \frac 12 (\ng-1)(\ng-2)\\[5pt]
\hline
\text{Total} & 2\ng &  \frac 1 2 \ng(\ng-1) & \frac 12 (\ng-1)(\ng-2)\\
\end{array}
\end{eqnarray*}
\caption{Parameters in the quark sector for $\ng$ generations. The $\Lambda_U$ and $\Lambda_D$ rows give the parameters if $Y_U$ or $Y_D$ are considered separately, and the third row gives the \emph{additional} parameters if both $Y_U$ and $Y_D$ are considered together. There are $(\ng-1)^2$ mixing parameters (angles plus phases), and a total of $(\ng^2+1)$ parameters.}
\end{table}

We choose a parameterization $V_{\text{CKM}} = \mathcal{V}(\theta_i, \delta_i )$ in terms of a standard functional form $\mathcal{V}$, where the $\ng(\ng-1)/2$ angles $\theta_i \in [0, \pi/2]$ and the $(\ng -1)(\ng-2)/2$ phases $\delta_i \in [0, 2 \pi)$. The CKM matrix for $\ng =3$ is given by~\cite{pdg}
\begin{eqnarray}\label{ckm}
&& \mathcal{V}(\theta_{12},\theta_{13},\theta_{23},\delta) \equiv 
\left[ \begin{array}{ccc} 
1 & 0 & 0 \\
0 & c_{23} & s_{23} \\
0 & -s_{23} & c_{23} \end{array} \right] \nn
&&\times
\left[ \begin{array}{ccc} 
c_{13} & 0 & s_{13}e^{-i \delta} \\
0 & 1 & 0 \\
- s_{13}e^{i \delta} & 0 & c_{13} \end{array} \right]
  \left[ \begin{array}{ccc} 
c_{12} & s_{12} & 0 \\
-s_{12} & c_{12} & 0\\
0 & 0 & 1 \end{array} \right]
\label{stdq}
\end{eqnarray}
where $s_i\equiv \sin \theta_i$ and $c_i\equiv \cos \theta_i$. It is now conventional to call the angles 
$\theta_{23},\theta_{13},\theta_{12}$ rather than $\theta_{1},\theta_2,\theta_3$. The standard form Eq.~(\ref{ckm}) has $\det \mathcal{V}=1$.

\medskip

\noindent \emph{Lepton Sector:}  The matrices $\Lambda_N$ and $\Lambda_E$ each have $\ng$ eigenvalues which are $CP$ even.  The lepton mixing matrices $V$ and $W$ are $\ng \times \ng$ unitary matrices, which can be parametrized by
\begin{eqnarray}
V &=& e^{i \chi}\ e^{i \Phi}\ \mathcal{V}(\theta_i, \delta_i) \ e^{i \Psi/2}, \ \nn 
W &=& e^{i \chi^\prime}\ e^{i \Phi^\prime}\ \mathcal{V}(\theta_i^\prime, \delta_i^\prime) 
\ e^{i \Psi^\prime/2}.
\end{eqnarray}
We use the same standard functional form $\mathcal{V}$ as for the quark sector, but with different numerical values for the arguments $\theta_i$ and $\delta_i$.\footnote{The use of the same symbols $\theta_i$ for the quark and lepton sectors should cause no confusion, since we do not need to deal with mixing in both sectors simultaneously.} The factor of two in $\Psi$ and $\Psi^\prime$ will be explained below.

The rephasing transformations $\Phi_\nu$, $\Phi_E$ and $\eta_N$ of Eq.~(\ref{rephasinghigh}) can be used to (i) eliminate $\chi$, $\chi^\prime$ and $\psi_i$, (ii) restrict $\psi_i^\prime$ to the range $[0, 2\pi)$ rather than $[0, 4 \pi)$, and (iii) eliminate \emph{either} $\Phi$ \emph{or} $\Phi^\prime$, \emph{but not both}.  It is convenient to use the same domain $[0, 2 \pi)$ for all phases, which is why $\Psi^\prime$ was scaled by a factor of $2$.

First consider amplitudes which depend only on $Y_\nu$ and $Y_E$, but not on $M$.  In this case, the mixing matrix $W$ is no longer observable and can be set to unity.  The mixing matrix $V$ has $(2 \ng - 1)$ allowed phase redefinitions: $n$ from  $\Phi_\nu$, $n$ from $\Phi_E$, and minus one, because $\Phi_\nu = \Phi_E \propto \openone$ does not change $V$.  Thus, the parameter counting for the mixing matrix $V$ is identical to that for $V_{\text{CKM}}$ in the quark sector, with $\ng(\ng-1)/2$ angles, and $(\ng-1)(\ng-2)/2$ phases. Similarly, for amplitudes depending only on $M$ and $Y_\nu$ and not on $Y_E$, the mixing matrix $V$ is no longer observable and can be set to unity.  The mixing matrix $W$ has $\ng$ allowed phase redefinitions $\Phi_\nu$.  Thus, there are $\ng(\ng-1)/2$ angles and $\ng(\ng+1)/2 -\ng=\ng(\ng-1)/2$ phases. If the three matrices $M$, $Y_\nu$ and $Y_E$ are considered together, then the mixing matrices $V$ and $W$ together can have $2 \ng$ allowed phase redefinitions due to $\Phi_\nu$ and $\Phi_E$. As compared with the case of only $V$ or only $W$, where there were $2\ng-1+\ng$ phase redefinitions possible, we have $(\ng-1)$ fewer phase redefinitions, and hence $(\ng-1)$ additional observable phases. These $(\ng-1)$ additional phases occur because the same phase redefinition $\Phi_\nu$ was present for both $V$ and $W$, and so cannot be chosen to remove phases from both $V$ and $W$. Thus, there are an additional $(\ng-1)$ phases if all three mass matrices are considered together. These phases can be included in either $V$ or $W$. The standard form of the mixing matrices which uses the $\Phi_\nu$ phases to eliminate the $\Phi$ phases from $V$ is given by
\begin{eqnarray}\label{vw1}
V &=& \mathcal{V}(\theta_i, \delta_i), \ \nn
W &=& e^{-i \bar \Phi} \ \mathcal{V}(\theta_i^\prime, \delta_i^\prime) \ e^{i \Psi^\prime/2} ,
\end{eqnarray}
whereas the standard form of the mixing matrices which uses the $\Phi_\nu$ phases to eliminate the $\Phi^\prime$ phases from $W$ is given by
\begin{eqnarray}\label{vw2}
V &=& e^{i \bar \Phi} \  \mathcal{V}(\theta_i, \delta_i), \ \nn
W &=& \mathcal{V}(\theta_i^\prime, \delta_i^\prime) \ e^{i \Psi^\prime/2} .
\end{eqnarray}
In Eq.~(\ref{vw1}), $V$ has the canonical CKM form with $\ng(\ng-1)/2$ angles $\theta_i$ and $(\ng -1)(\ng-2)/2$ phases $\delta_i$, whereas in Eq.~(\ref{vw2}), $W$ has the canonical PMNS form with $\ng(\ng-1)/2$ angles $\theta_i^\prime$ and $\ng(\ng-1)/2$ phases consisting of the $(\ng -1)(\ng -2)/2$ phases $\delta_i$ and the $(\ng -1)$ phases $\psi_i^\prime$.  In either basis, there are $(\ng -1)$ additional phases $\bar \Phi \equiv \Phi - \Phi^\prime$ which cannot be removed, and are observable. This parameter counting for $\mg = \ng$ is summarized in Table~II.

\begin{table}\label{tab:equal}
\begin{eqnarray*}
\begin{array}{c|ccc}
\text{Matrices} & \text{Masses} & \text{Angles} & \text{Phases} \\
\hline
\Lambda_N &  \ng & 0 & 0  \\
\Lambda_\nu &  \ng & 0 & 0  \\
\Lambda_E &  \ng & 0 & 0  \\
V: Y_\nu , Y_E  & 0 & \frac 1 2 \ng(\ng-1) &\frac 12 (\ng-1)(\ng-2)\\
W: M , Y_\nu  & 0 & \frac 1 2 \ng(\ng-1) &\frac 1 2 \ng(\ng-1)  \\
\bar \Phi \not \propto \openone   & 0 & 0 &\ng-1\\[5pt]
\hline
\text{Total} & 3\ng &  \ng(\ng-1) & \ng(\ng-1)\\
\end{array}
\end{eqnarray*}
\caption{Parameters in the lepton sector for $\mg = \ng$ generations. The $\Lambda_N$, $\Lambda_\nu$ and $\Lambda_E$ rows  give the parameters if $M$ or $Y_\nu$ or $Y_E$ are considered separately. The $V$ and $W$ rows give the \emph{additional} parameters if both $Y_\nu$ and $Y_E$, or both $M$ and $Y_\nu$ are considered together, respectively.  The last row gives the \emph{additional} parameters to those in the previous rows when all three matrices $M$, $Y_\nu$ and $Y_E$ are considered together. There are $2\ng(\ng-1)$ mixing parameters (angles and phases), and a total of 
$\ng(2\ng+1)$ parameters.}
\end{table}

\begin{table*}\label{tab:less}
\begin{eqnarray*}
\begin{array}{c|ccc}
\text{Matrices} & \text{Masses} & \text{Angles} & \text{Phases} \\
\hline
\Lambda_N &  \mg & 0 & 0  \\
\Lambda_\nu &  \mg  & 0 & 0  \\
\Lambda_E &  \ng & 0 & 0  \\
V: Y_\nu , Y_E  & 0 & \frac 1 2 \ng(\ng-1) &\frac 12 \ng (\ng-1)-\mg +1\\
W: M , Y_\nu  & 0 & \frac 1 2 \mg(\mg-1) &\frac 1 2 \mg(\mg-1) \\
\bar \Phi \not \propto \openone   & 0 & 0 &  \mg -1\\
U_{\ng - \mg } & 0 & \frac 1 2 (\ng - \mg) (\ng -\mg -1) &  
\frac 1 2 (\ng - \mg) (\ng -\mg +1) \\[5pt]
\hline
\text{Total} &  \ng + 2 \mg  & \ng \mg - \mg  & \ng \mg  - \ng  \\
\end{array}
\end{eqnarray*}
\caption{Parameters in the lepton sector for $\ng$ fermion generations and $\mg < \ng$ neutrino singlets.  The total number of parameters is equal to the sum of the first six rows minus the last row.  The parameters in $U_{\ng-\mg}$ are removed from $V$.}
\end{table*}

\begin{table*}\label{tab:more}
\begin{eqnarray*}
\begin{array}{c|ccc}
\text{Matrices} & \text{Masses} & \text{Angles} & \text{Phases} \\
\hline
\Lambda_N &  \mg & 0 & 0  \\
\Lambda_\nu & \ng & 0 & 0  \\
\Lambda_E &  \ng & 0 & 0  \\
V: Y_\nu , Y_E  & 0 & \frac 1 2 \ng(\ng-1) &\frac 12 (\ng-1) (\ng-2)\\
W: M , Y_\nu  & 0 & \frac 1 2 \mg(\mg-1) &\frac 1 2 \mg(\mg+1) -\ng \\
\bar \Phi \not \propto \openone   & 0 & 0 & \ng -1\\
U_{\mg - \ng } & 0 & \frac 1 2 (\mg - \ng) (\mg -\ng -1) &  
\frac 1 2 (\mg - \ng) (\mg -\ng +1) \\[5pt]
\hline
\text{Total} &  2 \ng + \mg   & \ng \mg - \ng  & \ng \mg  - \ng  \\
\end{array}
\end{eqnarray*}
\caption{Parameters in the lepton sector for $\ng$ fermion generations and $\mg > \ng$ neutrino singlets.  The total number of parameters is equal to the sum of the first six rows minus the last row.  The parameters in $U_{\mg -\ng}$ are removed from $W$.}
\end{table*}

\noindent \emph{$\vartheta$ Angles:} Once the mixing matrices have been put in standard form, one can perform additional phase rotations which leave the mixing matrices invariant to eliminate $\th$ angles. The only allowed transformation is an overall phase rotation with $\Phi_U=\Phi_D= \phi_Q\, \openone$, i.e.\ baryon number. Under this phase transformation,
\begin{eqnarray}
\th_3 &\to& \th_3,\nn
\th_2 &\to& \th_2 - 3\ng \phi_Q, \nn
\th_1 &\to& \th_1 + \frac32 \ng \phi_Q\ .
\end{eqnarray}
The transformation leaves $\th_3$ and $\th_2+2\th_1$ unchanged, so there are two physical $\th$ angles remaining: $\overline\th_{\text{QCD}}$, the strong interaction $CP$-angle in the basis where the quark mass matrices are real and diagonal, and $\overline \th_{EW}=\th_2+2\th_1$, the electroweak $CP$-angle in the basis where the quark and charged lepton mass matrices are real and diagonal.

\subsubsection{$\mg < \ng$}

For $\mg < \ng$, the $\mg \times \ng$ diagonal matrix $\Lambda_\nu$ can be written as
\begin{eqnarray}
\Lambda_\nu &\equiv& \left[ \begin{array}{cc}
\bar \Lambda_\nu & 0 \\
\end{array}\right],
\end{eqnarray}
where $0$ denotes the $\mg \times (\ng - \mg)$ zero matrix, and $\bar \Lambda_\nu$ is a diagonal $\mg \times \mg$ matrix with $\mg$ real non-negative eigenvalues. This matrix is invariant under
\begin{eqnarray}
\left[ \begin{array}{cc}
\bar \Lambda_\nu & 0 \\
\end{array}\right] &\to& e^{-i \Phi_\nu}\ \left[ \begin{array}{cc}
\bar \Lambda_\nu & 0 \\
\end{array}\right] \ \left[ \begin{array}{cc} 
e^{i \Phi_\nu} &  0 \\
0 & U_{\ng - \mg} \end{array}\right],\nn
\end{eqnarray}
where $U_{\ng - \mg}$ denotes an arbitrary $(\ng - \mg) \times (\ng - \mg)$ unitary matrix. The rephasing transformations of the lepton mixing matrices are
\begin{eqnarray}\label{rephasingless}
V &\to&  \left[ \begin{array}{cc} 
e^{-i \Phi_\nu} &  0 \\
0 & U^{-1}_{\ng - \mg} \end{array}\right]
\ V \ e^{i \Phi_E},\nn
W &\to& e^{-i \Phi_\nu}\ W\ \eta_N .
\end{eqnarray}
instead of Eq.~(\ref{rephasinghigh}).

The additional unitary transformation matrix in Eq.~(\ref{rephasingless}) can be used to eliminate parameters in $V$.  The parameter counting for $\mg < \ng$ is summarized in Table~III.  The number of $CP$-even parameters is $(\ng \mg + \ng + \mg)$ and the number of $CP$-odd parameters is $(\ng \mg - \ng)$, consistent with the results of Ref.~\cite{bgj1}.

\subsubsection{$\mg > \ng$}

For $\mg > \ng$, the $\mg \times \ng$ diagonal matrix $\Lambda_\nu$ can be written as
\begin{eqnarray}
\Lambda_\nu &\equiv& \left[ \begin{array}{c}
\bar \Lambda_\nu  \\
0 \\
\end{array}\right],
\end{eqnarray}
where $0$ denotes the $(\mg - \ng) \times \ng$ zero matrix, and $\bar \Lambda_\nu$ is a diagonal $\ng \times \ng$ matrix with $\ng$ real positive eigenvalues.  This matrix is invariant under
\begin{eqnarray}
\left[ \begin{array}{c}
\bar \Lambda_\nu  \\ 0 \\
\end{array}\right] 
&\to& \left[ \begin{array}{cc} 
e^{-i \Phi_\nu} &  0 \\
0 & U_{\mg - \ng} \end{array}\right]\ 
\left[ \begin{array}{c}
\bar \Lambda_\nu  \\ 0 \\ \end{array}\right] \ 
e^{i \Phi_\nu} , \nn
\end{eqnarray}
where $U_{\mg - \ng}$ denotes an arbitrary $(\mg - \ng) \times (\mg - \ng)$ unitary matrix. The rephasing transformation of the lepton mixing matrices is
\begin{eqnarray}\label{rephasingmore}
V &\to&  e^{-i \Phi_\nu} \ V \ e^{i \Phi_E},\nn
W &\to& \left[ \begin{array}{cc} 
e^{-i \Phi_\nu} &  0 \\
0 & U_{\mg - \ng} \end{array}\right]
\ W\ \eta_N .
\end{eqnarray}
instead of Eq.~(\ref{rephasinghigh}).

The additional unitary transformation matrix in Eq.~(\ref{rephasingmore}) can be used to eliminate parameters in $W$.  The parameter counting for $\mg > \ng$ is summarized in Table~IV.  The number of $CP$-even parameters is 
$(\ng \mg + \ng + \mg)$ and the number of $CP$-odd parameters is $(\ng \mg - \ng)$, consistent with the results of Ref.~\cite{bgj1}.

\subsection{Low-Energy Effective Theory}\label{subsec:leet}

The flavor matrices in the low-energy effective theory are written in Eq.~(\ref{low}) in the weak eigenstate basis.   These matrices are related to the mass eigenstate basis by
\begin{eqnarray}\label{translow}
Y_U &=& \Uc_{U^c} \ \Lambda_U \  \U_U , \nn
Y_D &=& \Uc_{D^c} \ \Lambda_D \ \U_D , \nn
Y_E &=& \Uc_{E^c} \ \Lambda_E \ \U_E , \nn
C_5 &=& {\Um}^{T}_{\nu}\ \Lambda_5 \ \Um_{\nu} .
\label{lowdiag}
\end{eqnarray}
Performing chiral flavor transformations in the low-energy theory with ${\calU_{U^c}}^T={\Uc_{U^c}}^{-1}$, ${\calU_{D^c}}^T={\Uc_{D^c}}^{-1}$, ${\calU_{E^c}}^T= {\Uc_{E^c}}^{-1}$, $\calU_Q={\U_U}^{-1}$, $\calU_L={\U_E}^{-1}$
brings the flavor matrices to the form
\begin{eqnarray}\label{diaglow}
Y_U &=&  \Lambda_U , \nn
Y_D &=& \Lambda_D \ V_{\text{CKM}}^{-1} , \nn
Y_E &=& \Lambda_E , \nn
C_5  &=& \left({U_{\text{PMNS}}^{-1}}\right)^T\  \Lambda_5 \ U_{\text{PMNS}}^{-1} ,
\end{eqnarray}
where $V_{\text{CKM}} \equiv \U_U {\U_D}^{-1}$ and $U_{\text{PMNS}}^{-1} \equiv \Um_\nu {\U_E}^{-1}$ are the two unitary matrices which describe flavor mixing in the low-energy effective theory.  $V_{\text{CKM}}$ is the CKM mixing matrix in the quark sector.  $U_{\text{PMNS}}$ is the PMNS mixing matrix in the lepton sector, which is the lepton mixing matrix which is physically measurable at low energies.

The real diagonal matrices $\Lambda_{U,D,E,5}$ are invariant under the rephasings
\begin{eqnarray}\label{rephasingfermionlow}
\Lambda_\psi &\to& e^{-i \Phi_\psi} \ \Lambda_\psi \ e^{i \Phi_\psi} ,\ \psi = U,D,E, \nn
\Lambda_5 &\to& \eta_\nu \ \Lambda_5 \ \eta_\nu ,
\end{eqnarray}
which correspond to arbitrary phase redefinitions of the fermion mass eigenstate fields $U^c$, $D^c$, $E^c$, $U$, $D$ and $E$, and the discrete rephasings $\nu \to \eta_\nu \nu$, where $\eta_\nu$ is a diagonal matrix with allowed eigenvalues $\pm 1$ for the low-energy Majorana neutrino fields.  Under these rephasings, the mixing matrices of the effective theory transform as
\begin{eqnarray}\label{rephasinglow}
V_{\text{CKM}} &\to& e^{-i \Phi_U}\ V_{\text{CKM}} \ e^{i \Phi_D},\nn
U_{\text{PMNS}} &\to& e^{-i \Phi_E}\ U_{\text{PMNS}}\ \eta_\nu\,.
\end{eqnarray} 

The quark mixing matrix $V_{\text{CKM}}$ has the angles and phases given in  Table~I as before.  The counting of parameters in the lepton sector is summarized in Table~V, and is well-known. $U_{\text{PMNS}}$ contains $\ng(\ng-1)/2$ angles $\theta_i$.  The number of phases of $U_{\text{PMNS}}$ is $\ng (\ng+1)/2$ minus the $\ng$ phase redefinitions $\Phi_E$, for a total of $\ng (\ng-1)/2$ phases consisting of $(\ng -1) (\ng -2)/2$ phases $\delta_i$ and $(\ng -1)$ phases  $\psi_i$.  The canonical parametrization of $U_{\text{PMNS}}$ is
\begin{eqnarray}\label{pmns}
U_{\text{PMNS}} &=& \mathcal{V}(\theta_i, \delta_i) \ e^{i \Psi/2} \ ,
\end{eqnarray}
$\Psi=\text{diag}(0,\psi_2,\ldots,\psi_n)$.

\begin{table}\label{tab:leet}
\begin{eqnarray*}
\begin{array}{c|ccc}
\text{Matrices} & \text{Masses} & \text{Angles} & \text{Phases} \\
\hline
\Lambda_E &  \ng & 0 & 0  \\
\Lambda_5 &  \ng & 0 & 0  \\
U_{\text{PMNS}} & 0 & \frac 1 2 \ng(\ng-1) &\frac 12 \ng (\ng-1)\\[5pt]
\hline
\text{Total} & 2 \ng &  \frac 1 2 \ng(\ng-1) & \frac 1 2 \ng(\ng-1)\\
\end{array}
\end{eqnarray*}
\caption{Parameters in the lepton sector of the low-energy effective theory for $\ng$ generations. The $\Lambda_E$ and $\Lambda_5$ rows  give the parameters if $m_E$ or $m_5$ are considered separately. The $U_{\text{PMNS}}$ row gives the mixing angles and phases of the PMNS mixing matrix.}
\end{table}

For $\ng =3$, the low-energy lepton mixing matrix is given by
\begin{eqnarray}
U_{\text{PMNS}} &=&
\mathcal{V} \left(\theta^{(U)}_{1},\theta^{(U)}_{2},\theta^{(U)}_{3},\delta^{(U)}\right)\nn
&& \times \left(\begin{array}{ccc}
1 & 0 & 0 \\
0 & e^{i \psi_{2}^{(U)}/2} & 0 \\
0 & 0 & e^{i \psi_{3}^{(U)}/2} \end{array}\right) \ ,
\label{u}
\end{eqnarray}
where the superscript $(U)$ denotes quantities in the PMNS matrix.

\section{Invariant Theory}\label{sec:invariants}

In the previous sections, we have discussed the parameters (masses, angles and phases) for the low- and high-energy theories. We would like to analyze the theories using  invariant quantities written directly in terms of the original parameters of the theory, the matrices $Y_{U,D,E,\nu}$ and $M$. The structure of the invariants is highly non-trivial, and depends in an interesting way on the number of generations.

To study the invariants, it is useful to introduce several mathematical results from invariant theory~\cite{goodmanwallach,procesi,derksen}. The general problem is the following: one has a set of variables $x_1,\ldots x_n$ which transform (reducibly or irreducibly) under the action of a group $G$. The set of polynomials in $\left\{x_i\right\}$ with complex coefficients form a ring $\mathbb{C}[x_1,\dots,x_n]$. The polynomial ring $\mathbb{C}[x_1,\dots,x_n]$ is a free ring on the generators $x_1,\ldots,x_n$, i.e.\ it is given by taking linear combinations of all possible products of powers of the generators with coefficients in $\mathbb{C}$, and there are no non-trivial relations among the generators.

The ring $\mathbb{C}[x_1,\dots,x_n]^G \subseteq \mathbb{C}[x_1,\dots,x_n] $ is the set of $G$-invariant polynomials, i.e.\ those polynomials which are unchanged by the action of $G$. This is clearly a ring, since sums and products of invariant polynomials are also invariant polynomials. A highly non-trivial result, if $G$ is a reductive group,\footnote{A reductive group is defined by the property that every representation is completely reducible. A Lie group which is a direct product of simple compact Lie groups and $U(1)$ factors is reductive, as is any finite group.} is that $\mathbb{C}[x_1,\dots,x_n]^G$ is finite generated.  Let the generators be $I_1,\ldots I_r$, each of which is a $G$-invariant polynomial in the original variables $x_1,\ldots, x_n$. Then, any $G$-invariant polynomial can be written as a polynomial $P \in  \mathbb{C}[I_1,\ldots,I_r]$. However, $\mathbb{C}[x_1,\dots,x_n]^G$ need not be a free ring in the generators $I_1,\ldots I_r$; there can be non-trivial relations among them. 

In the following sections, we analyze the invariant ring for the quark and lepton sectors of the Standard Model effective theory and the seesaw model. It is useful to first look at some simple examples before discussing the case of interest. We start with a famous result on symmetric polynomials, and then discuss three examples involving continuous groups which are closer in structure to the quark and lepton invariant problem. The first model is a theory which has a freely generated ring, with no relations. The second theory has one non-trivial relation, and is similar in structure to the ring for quark invariants for three generations studied in Sec.~\ref{sec:q2} and for lepton invariants in the Standard Model for two generations studied in Sec.~\ref{sec:sm2}. The third example is only slightly more complicated, but leads to an intricate structure of invariants, with many relations, and a complicated Hilbert series. This is similar to what we find for lepton invariants in the Standard Model for three generations, and in the seesaw model for two and three generations.

\subsection{Symmetric Polynomials}

The classic example from invariant theory is the study of symmetric polynomials. The permutation group $S_n$ acts on a  polynomial $f(x_1,\ldots,x_n)$ in $\mathbb{C}[x_1,\dots,x_n]$ by
\begin{eqnarray}
p: f(x_1,\ldots,x_n) \to  f(x_{p(1)},\ldots,x_{p(n)})
\end{eqnarray}
where $(p(1),\ldots,p(n))$ is a permutation of $(1,\ldots,n)$. A polynomial in $\mathbb{C}[x_1,\dots,x_n]^{S_n}$
is invariant under the action of any permutation. A standard result~\cite{weyl} is that the invariant ring is generated by the elementary symmetric polynomials
\begin{eqnarray}
I_1 &=& x_1 + x_2 + \ldots x_n = \sum_i x_i , \nn
I_2 &=& x_1 x_2 + x_1 x_3 + \ldots + x_{n-1}x_n = \sum_{i<j} x_ix_j , \nn
I_3 &=& x_1 x_2 x_3 +  \ldots + x_{n-2}x_{n-1} x_n = \sum_{i<j<k} x_ix_j x_k ,\nn
&\vdots &\nn
I_n &=& x_1 x_2 \ldots x_n .
\label{sympol}
\end{eqnarray}
In other words, any symmetric polynomial $f(x_1,\ldots,x_n)$ can be written as a \emph{polynomial} in $I_1,\ldots,I_n$, $f(x_1,\ldots,x_n)=g(I_1,\ldots,I_n)$, e.g.
\begin{eqnarray}
 x_1^2 + x_2^2 + \ldots x_n^2 &=& I_1^2-2I_2\,.
\end{eqnarray}
 The important point is that $g(I_1,\ldots,I_n)$ is a polynomial --- otherwise the result would be trivial, for knowing $I_1,\ldots ,I_n$, one could solve Eq.~(\ref{sympol}) to find $x_1,\ldots,x_n$, and hence determine $f$.

\subsection{Model I}\label{subsec:toy1}

Consider a theory with two couplings $m_1$ and $m_2$ which transform under a $G=U(1)\times U(1)$ symmetry as
\begin{eqnarray}\label{toym1m2}
m_1 \to  e^{i \phi_1} m_1,\qquad m_2 \to e^{i \phi_2} m_2 \ .
\end{eqnarray}
We look at the ring  $\mathbb{C}[m_1, m_1^*,m_2, m_2^*]^{U(1)\times U(1)}$ of all polynomials which are $U(1)\times U(1)$ invariant. It is clear that they can be written as linear combinations of monomials of the form
\begin{eqnarray}
\left( m_1 m_1^*\right )^{r_1} \left( m_2 m_2^*\right )^{r_2} 
\end{eqnarray}
where $r_1$ and $r_2$ are integers. Thus, the ring of invariant polynomials is generated by the invariants $I_1=m_1m_1^*$ and $I_2=m_2 m_2^*$, and there are no relations between these generators.

The Hilbert series $H(q)$ is defined as
\begin{eqnarray}
H(q) = \sum_{r=0}^{\infty} c_r q^r
\end{eqnarray}
where $c_r$ is the number of invariants of degree $r$, and $c_0=1$. In our example, $c_1=0$; $c_2=2$ since $m_1 m_1^*$ and $m_2 m_2^*$ are the two degree-two invariants; $c_3=0$; $c_4=3$ since $(m_1 m_1^*)^2$, $(m_1 m_1^*)(m_2 m_2^*)$ and $(m_2 m_2^*)^2$ are the three degree-four invariants; and so on. It is easy to see that the Hilbert series is
\begin{eqnarray}\label{hilbert1}
H(q) &=& 1 + 2 q^2 + 3 q^4 + 4 q^6 + 5 q^8 + \ldots\nn
&=& \sum_{n=0}^\infty (n+1) q^{2n}\nn
&=& \frac{1}{(1-q^2)^2}\ .
\end{eqnarray}

Another derivation of the Hilbert series is the following. The generators $I_1=m_1 m_1^*$ and $I_2=m_2 m_2^*$ are both of degree two, and the invariants of higher order are given by multiplying together arbitrary powers of $I_1$ and $I_2$.  The product
\begin{eqnarray}
\left(1 + I_1 + I_1^2 + \ldots\right)\left(1 + I_2 + I_2^2 + \ldots\right)
\end{eqnarray}
gives each invariant once, which leads to the Hilbert series
\begin{eqnarray}
H(q)&=&\left(1 +q^2 + q^4 + \ldots\right)\left(1 + q^2 + q^4 + \ldots\right)\nn
&=&\frac{1}{(1-q^2)^2},
\end{eqnarray}
in agreement with Eq.~(\ref{hilbert1}).

In the general case of a semisimple Lie group, it is known that $H(q)$ has the rational form
\begin{eqnarray}
H(q) = \frac{ N(q)}{D(q)},
\label{16}
\end{eqnarray}
where the numerator $N(q)$ and denominator $D(q)$ are polynomials. Furthermore, the numerator is of degree $d_N$ and is of the form
\begin{eqnarray}
N(q) = 1 + c_1 q  + \ldots c_{\dn-1} q^{\dn-1} + q^{\dn}
\end{eqnarray}
where the coefficients are non-negative, $c_r \ge 0$, and $N(q)$ is palindromic, i.e.
\begin{eqnarray}
N(q) = q^{\dn} N(1/q).
\end{eqnarray}
The denominator takes the form
\begin{eqnarray}
D(q) = \prod_{r=1}^p (1-q^{d_r}),
\end{eqnarray}
and is of degree $\dd=\sum_r d_r$.
The number of denominator factors $p$ is equal to the number of parameters. The number of parameters is defined as the minimal codimension of an orbit, and agrees with the usual physics usage of the term.
Model I has $p=2$ parameters, because we start with four objects $m_1$, $m_2$, $m_1^*$ and $m_2^*$ (or equivalently, the real and imaginary parts of $m_1$ and $m_2$), and have two phase redefinitions Eq.~(\ref{toym1m2}), which eliminates two variables.  In other words, one can always make a phase redefinition to make $m_1$ and $m_2$ real and non-negative, and these are the two independent parameters. In our example, $N(q)=1$, $d_1=d_2=2$ and the number of denominator factors is two.  The number of denominator factors $p$ is equal to the number of parameters.

There is a theorem due to Knop~\cite{knop} which says that
\begin{eqnarray}\label{knopthm}
\dim V \ge \dd - \dn \ge p
\end{eqnarray}
where $\dim V$ is the dimension of the vector space on which the group transformations act; $\dd$ and $\dn$ are the degrees of the  denominator and numerator; and $p$ is the number of parameters. In most cases, the upper bound is an equality, but not always. (We will see an example for the quark invariants involving only the $U$-quark mass matrix.) In Model I, the vector space basis is $m_1$, $m_1^*$, $m_2$, $m_2^*$, so $\dim V=4$, $p=2$, $\dn=0$ and $\dd=\sum d_r =4$, and we see that Knop's theorem gives $4 \ge 4-0 \ge 2$, with an equality for the upper bound.

One also can construct a multi-graded Hilbert series. Let $c_{r_1r_2r_3 r_4}$ be the number of invariants of order $r_1$ in $m_1$, order $r_2$ in $m_1^*$, order $r_3$ in $m_2$, and order $r_4$ in  $m_2^*$. Then
\begin{eqnarray}
h(q_1,q_2,q_3,q_4) &=& \sum c_{r_1r_2r_3r_4} q_1^{r_1}q_2^{r_2}q_3^{r_3}q_4^{r_4}\nn
&=& \frac{1}{(1-q_1q_2)(1-q_3q_4)},
\end{eqnarray}
and the usual Hilbert series is $H(q)=h(q,q,q,q)$. The multi-graded series gives more information about the structure of the invariants. However, it is important to remember that the results quoted above for $H(q)$, Eqs.~(\ref{16})--(\ref{knopthm}), do not hold in general for the multi-graded case.

\subsection{Model II}\label{subsec:toy2}

Consider a theory with couplings $m_1$ and $m_2$ with charges one and two, respectively, under a $G=U(1)$ symmetry,
\begin{eqnarray}\label{toy2}
m_1 \to  e^{i \phi} m_1,\qquad m_2 \to e^{2i \phi} m_2 \ .
\end{eqnarray}
The ring of invariant polynomials $\mathbb{C}[m_1,m_1^*,m_2,m_2^*]^{U(1)}$ is generated by the four
basic invariants $I_1=m_1m_1^*$, $I_2=m_2m_2^*$, $I_3=m_2m_1^{*2}$ and $I_4=m_2^* m_1^2$. These generators, however, are not all independent, since $I_3 I_4 =I_1^2 I_2 $, so that $\mathbb{C}[m_1,m_1^*,m_2,m_2^*]^{U(1)}$ is not a free ring generated by $I_{1-4}$.

It is straightforward to show that the multi-graded Hilbert series is
\begin{eqnarray}\label{hilberttoy2}
h(q_1,q_2,q_3,q_4) &=& \frac{1-q_1^2 q_2^2q_3q_4}{(1-q_1q_2)(1-q_3q_4)(1-q_3q_2^2)(1-q_4q_1^2)},\nn
\end{eqnarray}
where $q_1$, $q_2$, $q_3$ and $q_4$ count powers of $m_1$, $m_1^*$, $m_2$ and $m_2^*$, respectively.

The denominator of the multi-graded Hilbert series is generated by the invariants $I_{1-4}$, whereas the numerator compensates for the fact that $I_3 I_4$ and $I_1^2 I_2 $ count as only one invariant at order $q_1^2 q_2^2 q_3 q_4$, rather than two, because $I_3 I_4 =I_1^2 I_2 $. The numerator of the multi-graded Hilbert series does not have the special properties of the numerator of the Hilbert series $H(q)$ discussed in the previous example.  

In this example, $\dim V=4$, $\dim G=1$, and there are three parameters since the phase transformation Eq.~(\ref{toy2}) eliminates one of the original four real variables in $m_1$ and $m_2$.
The Hilbert series $H(q)=h(q,q,q,q)$ is
\begin{eqnarray}\label{Hilberttoy2}
H(q) &=& \frac{1+q^3}{(1-q^2)^2(1-q^3)},
\end{eqnarray}
which has a palindromic numerator with $\dn=3$, and a denominator with $d_D=7$, and $p=3$ is equal to the number of denominator factors and to the number of parameters.  Knop's theorem gives $4 \ge 7-3 \ge 3$, with an equality for the upper bound.

Expanding Eq.~(\ref{Hilberttoy2}) in a series in $q$ gives the invariants of each degree. We see that there are two generators of degree two, $I_1$ and $I_2$, and one generator of degree three, which can be chosen to be $I_3+I_4$, corresponding to the denominator factors $(1-q^2)^2$ and $(1-q^3)$, respectively. Expanding out the denominator would give a coefficient of $q^3$ of $+1$. There are two invariants of degree three, $I_3 \pm I_4$.
The missing degree-three invariant $I_3-I_4$ is counted by the $+q^3$ term in the numerator, so that the coefficient of $q^3$ in the expansion of $H(q)$ is $2$. When the denominator factors are expanded in a series, they can occur to any power, so one can have arbitrary powers of $I_1$, $I_2$ and $I_3+I_4$. However, the $q^3$ factor in the numerator occurs only once. This means that powers of $I_3-I_4$ higher than the first can all be eliminated in terms of polynomials $P(I_1,I_2,I_3+I_4)$ which have already been included. This statement follows from the identity
\begin{eqnarray}
(I_3-I_4)^2 &=& (I_3+I_4)^2-4 I_3 I_4 \nn
&=& (I_3+I_4)^2-4I_1^2 I_2.
\label{eq59}
\end{eqnarray}
There exists a similar identity for the Jarlskog invariant which will be derived in Sec.~\ref{sec:quarkinvariants}.

The generator $I_3+I_4$ of the denominator is not homogeneous in the multi-grading; $I_3$ is of degree $q_3q_2^2$ and $I_4$ is of degree $q_4 q_1^2$, which is why Eq.~(\ref{hilberttoy2}) can not be written in a form similar to Eq.~(\ref{Hilberttoy2}) with positive coeficients in the numerator and one less generator in the denominator.

\subsection{Model III}\label{subsec:toy3}

Consider yet another model with three couplings $m_1$, $m_2$ and $m_3$ with charges $1$, $2$ and $3$, respectively, under a $U(1)$ symmetry,
\begin{eqnarray}\label{toy3}
m_1 \to e^{i \phi}m_1,\ m_2 \to e^{2i\phi} m_2,\ m_3 \to e^{3i\phi} m_3 \ .
\end{eqnarray}
The structure of the invariants is considerably more complicated than in the previous examples, even though the theory is only slightly more complicated. All the invariant polynomials are generated by thirteen invariant generators
\begin{eqnarray}
I_1 &=& m_1 m_1^*,\nn
I_2 &=& m_2 m_2^*,\nn
I_3 &=& m_3 m_3^*,\nn
I_4 &=& m_1^2m_2^*,\nn
I_5&=&  m_1^{*2} m_2,\nn
I_6&=& m_1^3 m_3^*,\nn
I_7&=& m_1^{*3} m_3,\nn
I_8 &=& m_2^3 m_3^{*2},\nn
I_9 &=& m_2^{*3} m_3^2,\nn
I_{10} &=& m_1 m_2 m_3^*,\nn
I_{11} &=& m_1^* m_2^* m_3,\nn
I_{12} &=& m_1 m_3 m_2^{*2},\nn
I_{13} &=& m_1^* m_3^*  m_2^2 .
\label{eq61}
\end{eqnarray}
There are 35 relations between products of invariants $I_i I_j$ given by: $I_4 I_5 =I_1^2 I_2$, $I_4 I_7 = I_1^2 I_{11}$, $I_4 I_8 = I_2 I_{10}^2$, $I_4 I_9 = I_{12}^2$,  $I_4 I_{10} = I_2 I_6$, $I_4 I_{11} = I_1 I_{12}$, $I_4 I_{13} = I_1 I_2 I_{10}$, 
$I_5 I_6 = I_1^2 I_{10}$, $I_5 I_8 = I_{13}^2$, $I_{5} I_9 = I_2 I_{11}^2$, 
$I_5 I_{10} = I_1 I_{13}$, $I_5 I_{11} = I_2 I_7$, $I_5 I_{12} = I_1 I_2 I_{11}$, $I_6 I_7 = I_1^3 I_3$, $I_6 I_8 = I_{10}^3$, $I_6 I_9 = I_3 I_4 I_{12}$, $I_6 I_{11} = I_1 I_3 I_4$, $I_6 I_{12} = I_3 I_4^2$, $I_6 I_{13} = I_1 I_{10}^2$, $I_7 I_8 = I_3 I_5 I_{13} $, $I_7 I_9 = I_{11}^3$, $I_7 I_{10} = I_1 I_3 I_5$, $I_7 I_{12} = I_1 I_{11}^2$, $I_7 I_{13} = I_3 I_5^2$, $I_8 I_9 = I_2^3 I_3^2$, $I_8 I_{11} = I_2 I_3 I_{13}$, $I_8 I_{12} = I_2^2 I_3 I_{10}$, $I_9 I_{10} = I_2 I_3 I_{12}$, $I_9 I_{13} = I_2^2 I_3 I_{11}$, $I_{10} I_{11} = I_1 I_2 I_3$, $I_{10} I_{12} = I_2 I_3 I_4$, $I_{10} I_{13} = I_1 I_8$, $I_{11} I_{12} = I_2 I_3 I_5$, $I_{11} I_{13} = I_2 I_3 I_5$ and $I_{12} I_{13} = I_1 I_2^2 I_3$. The new feature here is that these relations are not independent---there are relations among the relations (known as syzygies in the mathematics literature), e.g.\ multiplying both sides of $I_4I_7 = I_1^2I_{11}$ and $I_5 I_6 = I_1^2 I_{10}$ gives
\begin{eqnarray}
I_4 I_5 I_6 I_7 &=& I_1^4 I_{10} I_{11} \ ,
\end{eqnarray}
which is also obtained by multiplying the relations $I_4 I_5=I_1^2 I_2$ and $I_6I_7=I_1^3I_3$, and using $I_{10} I_{11} = I_1 I_2 I_3$. The Hilbert series is
\begin{eqnarray}\label{hilberttoy3}
H(q)&=&\frac{1+q^2+3q^3+4q^4+4q^5+4q^6+3q^7+q^8+q^{10}}{(1-q^2)^2
(1-q^3)(1-q^4)(1-q^5)} \ .\nn
\label{eq63}
\end{eqnarray}
Here $\dim V=6$, $\dim G=1$, and the number of parameters is $5$. From the Hilbert series,
$d_N=10$, $d_D=16$, and $p=5$. The number of parameters is equal to $p$, and Knop's theorem gives $6 \ge 16-10 \ge 5$, with an equality for the upper bound.

There are thirteen invariants in Eq.~(\ref{eq61}). However, there are only five denominator factors in Eq.~(\ref{eq63}), so only five basic invariants, two of degree two, and one each of degrees three,  four and five, generate a free ring. The other invariants must satisfy non-trivial relations (those given below Eq.~(\ref{eq61})), and this is reflected by the complicated numerator in Eq.~(\ref{eq63}), which implies  that the invariant ring has a non-trivial structure, with many relations. The different terms in the numerator show that there are many invariants which can be eliminated when raised to higher powers, or multiplied by lower order invariants, by relations analogous to Eq.~(\ref{eq59}). There is one invariant of degree two (the $+q^2$ term), three in degree three (the $+3q^3$ term), etc. This model shows that even a relatively simple theory given by Eq.~(\ref{toy3}) can lead to a set of invariants with an interesting syzygy structure. Furthermore, the number of invariants and relations of each degree is encoded in the Hilbert series.

\section{Quark Invariants}\label{sec:quarkinvariants}

We can now address the first problem of interest --- flavor invariants in the quark sector.  We are interested in polynomials in $m_U$, ${m_U}^\dagger$, $m_D$ and ${m_D}^\dagger$ where
\begin{eqnarray}\label{mumd}
m_U &\to& {\calU_{U^c}}^{T} \ m_U \ \calU_Q , \nn
m_D &\to& {\calU_{D^c}}^T \ m_D \ \calU_Q ,
\label{qtrans}
\end{eqnarray}
under the chiral flavor transformations.\footnote{One could equally well work with the Yukawa matrices, which differ by  factor $v/\sqrt 2$.} To cancel $\calU_{U^c}$ and $\calU_{D^c}$, one must consider the combinations
\begin{eqnarray}\label{xuxd}
X_{U} &\equiv& {m_{U}}^\dagger m_{U} , \nn
X_{D} &\equiv& {m_{D}}^\dagger m_{D} , 
\end{eqnarray}
which both transform as adjoints
\begin{eqnarray}\label{transxud}
X_{U,D} &\to& \calU_Q^{\dagger} \ X_{U,D} \ \calU_Q.
\end{eqnarray}
Thus, the invariants are traces of products of $X_{U}$ and $X_{D}$. The structure of the invariants depends non-trivially on the number of generations, so we consider the cases $\ng=2$ and $\ng=3$ separately.

\subsection{$\ng=2$}

First, consider invariants involving only $X_U$. The basic invariants are
\begin{eqnarray}
\tr{X_U},\tr{{X_U}^2 },\tr{{X_U}^3 },\ldots
\end{eqnarray}
where $\tr{*}$ denotes a matrix trace.
This series of traces terminates after $\ng$ terms for an $\ng \times \ng$ matrix, by the Cayley-Hamilton theorem which states that every matrix satisfies its characteristic equation. For an arbitrary $2 \times 2$ matrix $A$, the Cayley-Hamilton theorem gives
\begin{eqnarray}\label{a2}
A^2&=  \Tr{A}A +  \frac12 \ \left[\Tr{A^2}- \Tr{A}^2\right]\openone.
\end{eqnarray}
Taking the trace of both sides gives the trivial result $\vev{A^2}=\vev{A^2}$.
Multiplying by $A$ and taking the trace implies
\begin{eqnarray}\label{tra3}
\Tr{A^3} &=\frac32\Tr{A}\Tr{A^2}-\frac12\Tr{A}^3,
\end{eqnarray}
so that $\tr{A^n}$, $n\ge 3$ can be written in terms of $\tr{A}$ and $\tr{A^2}$. Thus, there are two independent invariants, $I_{2,0}=\tr{X_U}$ and $I_{4,0}=\tr{{X_U}^2 }$, which can be constructed from $X_U$ alone. Both of these invariants are $CP$ even. The two invariants contain the same information as the eigenvalues of $X_U$, i.e.\ the two $U$-type quark masses.  For invariants constructed only from $m_U$, the number of parameters is $p=2$, the two eigenvalues of $X_U$. The vector space has $\dim V=8$, because $m_U$ and ${m_U}^\dagger$ are both $2\times 2$ matrices, and $I_{2,0}$ and $I_{4,0}$ are of degree two and four, respectively, in $m_U$, so the Hilbert series is
\begin{eqnarray}\label{Hilbertmu}
H(q) &=& \frac{1}{(1-q^2)(1-q^4)}.
\end{eqnarray}
Here $\dn=0$, $\dd=6$ are the degrees of the numerator and denominator, respectively, and the number of denominator factors is $p=2$, which is equal to the number of parameters.  Knop's theorem gives $8 \ge 6-0 \ge 2$, which holds, but this time the upper bound is not an equality.

Similarly, there are two independent $CP$-even invariants $I_{0,2}=\tr{X_D}$ and $I_{0,4}=\tr{{X_D}^2 }$ which involve only $X_D$.  These two invariants contain the same information as the eigenvalues of $X_D$, namely the two $D$-type quark masses.

Invariants containing both $X_U$ and $X_D$ can be written as traces of the form
\begin{eqnarray}\label{xuxdgeneral}
\tr{ {X_U}^{r_1} {X_D}^{s_1}{X_U}^{r_2} {X_D}^{s_2}\ldots},
\end{eqnarray}
for integers $r_i$ and $s_i$.
The Cayley-Hamilton theorem for a $2 \times 2$ matrix, Eq.~(\ref{a2}), implies that all powers $r_i$ and $s_i$ greater than one in Eq.~(\ref{xuxdgeneral}) can be reduced, so we are left with traces of the form
\begin{eqnarray}
\tr{ X_U X_D \ldots  X_U X_D}=\tr{ (X_U X_D)^r}.
\end{eqnarray}
Again, invariants with $r>1$ can be rewritten in terms of lower order invariants, so there is only one independent invariant, $I_{2,2}=\tr{ X_U X_D }$, which is $CP$ even.

In summary, the basic quark invariants for $\ng=2$ quark generations, which generate all the invariants, are:
\begin{eqnarray}\label{quarkinvs}
I_{2,0} &=& \tr{X_U} = \tr{ {m_U}^\dagger m_U}, \nn
I_{0,2} &=& \tr{X_D}= \tr{ {m_D}^\dagger m_D}, \nn
I_{4,0} &=& \tr{{X_U}^2} = \tr{ \left({m_U}^\dagger m_U  \right)^2 }, \nn
I_{2,2} &=& \tr{X_U X_D} = \tr{ {m_U}^\dagger m_U  {m_D}^\dagger m_D},\nn
I_{0,4} &=& \tr{{X_D}^2}= \tr{ \left({m_D}^\dagger m_D\right)^2 }.
\label{eq77}
\end{eqnarray}
Writing the invariants in terms of the usual quark masses and the Cabibbo angle gives
\begin{eqnarray}
I_{2,0} &=& m_u^2+m_c^2,\nn
I_{0,2} &=& m_d^2+m_s^2,\nn
I_{4,0} &=& m_u^4+m_c^4,\nn
I_{2,2} &=& m_u^2 m_s^2+m_c^2m_d^2+ (m_s^2-m_d^2)(m_c^2-m_u^2)\cos^2\theta ,\nn
I_{0,4} &=& m_d^4+m_s^4.
\end{eqnarray}
Knowing the five invariants allows one to determine the four masses and $\theta$, because $m_i \ge 0$, and $\theta$ lies in the first quadrant.

Using $u$ and $d$ to count powers of $m_U$ and $m_D$ gives the multi-graded Hilbert series
\begin{eqnarray}\label{hilbertud}
h(u,d) = \frac{1}{(1-u^2)(1-u^4)(1-d^2)(1-d^4)(1-u^2d^2)}.\nn
\end{eqnarray}
The Hilbert series $H(q)=h(q,q)$ is
\begin{eqnarray}\label{Hilbertq}
H(q)= \frac{1}{(1-q^2)^2(1-q^4)^3}.
\label{hilb1}
\end{eqnarray}
In this example, $p=5$ (four masses and one mixing angle, see Table~I), $\dim V=16$, since there are four $2 \times 2$ matrices, $\dn=0$, and $\dd=16$. The number of denominator factors is the number of parameters, and Knop's theorem gives $16 \ge 16-0 \ge 5$, with the upper bound an equality.

The denominator factors in Eq.~(\ref{hilb1}) show that there are  two generators of degree two, and three of degree four, which agrees with Eq.~(\ref{eq77}).

If one started with $X_U$ and $X_D$ as the basic objects, then $\dim V = 8$. In this case, the Hilbert series is given by replacing $q^2 \to q$ in Eq.~(\ref{hilb1}), since we now count powers of $X_U,X_D$ rather than $m_U,m_D$, so $\dn=0$, $\dd=8$
and Knop's inequality becomes $8 \ge 8 - 0 \ge 5$.

\subsection{$\ng=3$}\label{sec:q2}

For an arbitrary $3 \times 3$ matrix $A$, the Cayley-Hamilton theorem states that
\begin{eqnarray}\label{a3id}
A^3 &=&A^2 \Tr A - \frac12A \left[\Tr{ A}^2-\Tr {A^2}\right]\nn
&&+\frac 1 6\left[\Tr{ A}^3-3 \Tr{ A^2} \Tr A + 2 \Tr {A^3}\right]\openone.
\end{eqnarray}
Taking the trace of both sides gives the trivial result $\vev{A^3}=\vev{A^3}$.
Multiplying by $A$ and taking the trace gives
\begin{eqnarray}\label{tra4id}
\Tr{ A^4} &=& \frac1{6}\Tr{ A}^4 -  \Tr {A }^2 \Tr{ A^2}
+ \frac 4 3 \Tr {A^3} \Tr A  + \frac 1 2 \Tr{A^2}^2,\nn
\end{eqnarray}
so that $\tr{A^n}$, $n\ge 4$ can be rewritten in terms of $\tr{A}$, $\tr{A^2}$, and $\tr{A^3}$.

Thus, the invariants involving $X_U$ alone are $I_{2,0}=\tr{X_U}$, $I_{4,0}=\tr{{X_U}^2 }$ and $I_{6,0}=\tr{{X_U}^3 }$, and invariants involving $X_D$ alone are $I_{0,2}=\tr{X_D}$, $I_{0,4}=\tr{{X_D}^2 }$ and $I_{0,6}=\tr{{X_D}^3 }$, all of which are $CP$ even.

Invariants containing both $X_U$ and $X_D$ are of the form Eq.~(\ref{xuxdgeneral}), but now with $r_i=1,2$ and $s_i=1,2$, so that one has traces of products of $X_U,X_U^2,X_D,X_D^2$. This restriction still leads to an infinite number of invariants.  However, many of these invariants are not independent.  For arbitrary $3 \times 3$ matrices $A$, $B$ and $C$, one has the identity
\begin{eqnarray}
0 &=& \Tr{A}^2 \Tr B \Tr C - \Tr{BC} \Tr{A}^2-2 \Tr{AB} \Tr A \Tr C\nn
&&- 2 \Tr{AC} \Tr A \Tr B+2 \Tr{ABC} \Tr A +2 \Tr{ACB} \Tr A\nn
&&- \Tr{A^2} \Tr B \Tr C +2 \Tr{AB} \Tr{AC} + \Tr{A^2} \Tr{BC}\nn
&& +2 \Tr{C} \Tr{A^2B}+2 \Tr{B}\Tr{A^2C}-2\Tr{A^2BC}\nn
&&-2\Tr{A^2CB}-2\Tr{ABAC} 
\label{eq83}
\end{eqnarray}
which can be derived by substituting $A \to A + B+C$ into Eq.~(\ref{tra4id}), and picking out the order $A^2BC$ terms. This identity eliminates $\vev{ABAC}$, i.e.\ traces where the same matrix is repeated, so that in invariants Eq.~(\ref{xuxdgeneral}), $X_U$, ${X_U}^2$, $X_D$ and ${X_D}^2$ can each occur at most once. For example, $\vev{X_U \ldots X_U \ldots}$ can be replaced by $\vev{X_U^2 \ldots}$, and $\vev{X_U^2 \ldots X_U^2 \ldots}$ can be replaced by $\vev{X_U^4 \ldots}$, which can then be eliminated using Eq.~(\ref{a3id}).

Writing out all of the possibilities gives the basic quark invariants for $\ng=3$ quark generations. There are 11 $CP$-even invariants, ten of which are
\begin{eqnarray}\label{xudinvariantse}
I_{2,0}&=& \tr{ X_U},\nn
I_{0,2}&=& \tr{ X_D},\nn
I_{4,0}&=& \tr{ {X_U}^2},\nn
I_{2,2}&=& \tr{ X_U  X_D},\nn
I_{0,4}&=& \tr{ {X_D}^2},\nn
I_{6,0}&=& \tr{ {X_U}^3},\nn
I_{4,2}&=& \tr{ {X_U}^2  X_D},\nn
I_{2,4}&=& \tr{   X_U {X_D}^2},\nn
I_{0,6}&=& \tr{ {X_D}^3},\nn
I_{4,4} &=& \tr{ {X_U}^2 {X_D}^2},
\label{eq84}
\end{eqnarray}
and one $CP$-odd invariant
\begin{eqnarray}\label{xudinvariantso}
I_{6,6}^{(-)} &=& \tr{ {X_U}^2 {X_D}^2  X_U  X_D} -  \tr{{X_D}^2 {X_U}^2 X_D  X_U }\,.\nn
\label{eq85}
\end{eqnarray}
The eleventh $CP$-even invariant is 
\begin{eqnarray}
I_{6,6}^{(+)} &=& \tr{ {X_U}^2 {X_D}^2  X_U  X_D} +  \tr{{X_D}^2 {X_U}^2 X_D  X_U}\,.\nn
\label{eq86}
\end{eqnarray}
All the invariants in the quark sector can be written as polynomials in these basic invariants.

The multi-graded and one-variable Hilbert series are
\begin{widetext}
\begin{eqnarray}
h(u,d) &=& \frac{1+u^6d^6}{(1-u^2)(1-u^4)(1-u^6)(1-d^2)(1-d^4)(1-d^6)(1-u^2d^2)(1-u^4d^2)(1-u^2d^4)(1-u^4d^4)},\nn
H(q)&=&h(q,q)=\frac{1+q^{12}}{(1-q^2)^2(1-q^4)^3(1-q^6)^4(1-q^8)},
\label{hilb2}
\end{eqnarray}
\end{widetext}
respectively.  This case has $p=10$ parameters, consisting of 6 masses, three angles and one phase, which agrees with the number of denominator factors. The original variable space has $\dim V=36$, from the two $3 \times 3$ mass matrices and their complex conjugates. The degrees of the numerator and denominator are $\dn=12$ and $\dd=48$, respectively, and Knop's inequality is $36 \ge 48-12 \ge 10$, which is satisfied, with the upper bound being an equality. If one started with $X_U$ and $X_D$ as the basic objects, then $\dim V = 18$, and the Hilbert series is given by replacing $q^2 \to q$ in Eq.~(\ref{hilb2}), so $\dn=6$, $\dd=24$, and Knop's inequality becomes $18 \ge 24 - 6 \ge 10$.

The denominator of Eq.~(\ref{hilb2}) shows that there are two invariants of degree two,  three of degree four, four of degree six,  and one of degree eight, which can occur multiplied in arbitrary combinations, with no relations among them. This is a total of 10 invariants, which are the ones listed in Eq.~(\ref{eq84}). One can see that their degrees match the denominator factors in Eq.~(\ref{hilb2}). What about the remaining two invariants? The numerator factor of Eq.~(\ref{hilb2}) shows that there is one additional invariant of degree twelve other than those given by products of denominator factors. This is the $CP$-odd invariant Eq.~(\ref{eq85}). The Hilbert series implies that the other degree-twelve invariant, Eq.~(\ref{eq86}), cannot be an independent invariant. Indeed, it can be written as a polynomial in the other $CP$-even invariants, 
\begin{eqnarray}
   3 I_{6,6}^{(+)}&=&
   I_{2,0}^3 I_{0,2}^3-I_{2,0} I_{4,0} I_{0,2}^3-3
   I_{2,2} I_{2,0}^2 I_{0,2}^2\nn
   &&+3 I_{4,2}
   I_{2,0} I_{0,2}^2
   -I_{0,4} I_{2,0}^3 I_{0,2}+3
   I_{2,4} I_{2,0}^2 I_{0,2}\nn
   &&-3 I_{4,4} I_{2,0}
   I_{0,2}\nn
   &&+I_{0,4} I_{6,0} I_{0,2}+3 I_{2,4}
   I_{4,2}+3 I_{2,2} I_{4,4}\nn
&&   +I_{0,6} I_{2,0}
   I_{4,0}-I_{0,6} I_{6,0},
   \label{eq88}
\end{eqnarray}
and so can be eliminated.

The Hilbert series numerator only has an entry $q^{12}$, but there is no $q^{24}$ term. This means that $I_{6,6}^{(-)}$ is an independent invariant, but the square and all higher powers of $I_{6,6}^{(-)}$  are not. The square of the $CP$-odd invariant $I_{6,6}^{(-)}$ is $CP$-even, and can be written as a polynomial (with 241 terms out of a possible 305 terms) in the $CP$-even invariants
of Eq.~(\ref{eq84}). The most general polynomial invariant in the quark sector can be written as
\begin{eqnarray}
P_1+I_{6,6}^{(-)}\ P_2
\end{eqnarray}
where $P_1$ and $P_2$ are polynomials in the $CP$-even invariants Eq.~(\ref{xudinvariantse}). 

This example illustrates how the structure of the invariants is encoded in the Hilbert series. For many purposes, the details of the relations, such as Eq.~(\ref{eq88}), or the formula for $\left(I_{6,6}^{(-)}\right)^2$ are not important; all one needs to know is that $I_{6,6}^{(-)}$ occurs linearly, and $I_{6,6}^{(+)}$ can be eliminated.

The quark sector parameters are determined by the ten $CP$-even parameters $I_{2,0}$, $I_{4,0}$, $I_{6,0}$, $I_{0,2}$, $I_{0,4}$, $I_{0,6}$, $I_{2,2}$, $I_{2,4}$, $I_{4,2}$, $I_{4,4}$, and the single $CP$-odd parameter $I_{6,6}^{(-)}$.  From the $CP$-even invariants, one can determine the $U$-type quark masses $m_{u,c,t}$ and $D$-type quark masses $m_{d,s,b}$, which are real and non-negative, and four combinations of the CKM parameters, $\cos\theta_{12}$, $\cos\theta_{13}$, $\cos\theta_{23}$ and $\cos \delta$, all of which are $CP$ even. Since the CKM angles $\theta_{12}$, $\theta_{13}$, $\theta_{23}$ lie in the first quadrant, these angles are determined uniquely by their cosines.  However, $\cos \delta$ does not determine the phase $\delta$ uniquely, because it cannot distinguish between $\delta$ and $-\delta$.  Under $CP$, $\delta \leftrightarrow -\delta$. Thus, one $Z_2$ piece of information, the sign of $\delta$, is missing. This sign is provided by the invariant $I_{6,6}^{(-)}$. The only information needed is the sign of $I_{6,6}^{(-)}$, which is why $\left({I_{6,6}^{(-)}}\right)^2$ can be written in terms of the other $CP$-even invariants. This discussion corresponds to the well-known result that the unitarity triangle can be obtained by measuring the lengths of its sides, which are $CP$-conserving, rather than the angles, which are $CP$-violating. Knowing the sides determines the triangle up to a two-fold reflection ambiguity, which is fixed by the sign of $I_{6,6}^{(-)}$, or, equivalently, the sign of the Jarlskog invariant, so that the only additional information contained in the Jarlskog invariant is the sign. The relations between the invariants are similar to those obtained by studying rephasing invariants~\cite{jmrephasing}. 

The invariant  $I_{6,6}^{(-)}$ also can be written as
\begin{eqnarray}
 I_{6,6}^{(-)} &=& \frac13\Tr { \left[ X_U, X_D \right]^3}\ ,
 \end{eqnarray}
and is proportional to the Jarlskog invariant $J$~\cite{jarlskog},
\begin{eqnarray}
 I_{6,6}^{(-)} &=& 2 i J (m_c^2-m_u^2)(m_t^2-m_c^2)(m_t^2-m_u^2)\nn
&&\times (m_s^2-m_d^2)(m_b^2-m_s^2)(m_b^2-m_d^2) ,
\end{eqnarray}
where
\begin{eqnarray}
J  = \text{Im} \, \left(V_{\text{CKM}}\right)_{11}\left(V_{\text{CKM}}\right)_{12}^*\left(V_{\text{CKM}}\right)_{22} \left(V_{\text{CKM}}\right)_{21}^*\,.
\end{eqnarray}  
$ I_{6,6}^{(-)}$ vanishes if two $U$-type quarks or two $D$-type quarks are degenerate. It is well-known that quark $CP$ violation vanishes for degenerate $U$-type or $D$-type quarks. $ I_{6,6}^{(-)}$ is odd under the exchange of two $U$-type or two $D$-type masses, e.g under $m_u \leftrightarrow m_c$, whereas the invariants in Eq.~(\ref{eq84}) are even under exchange, so 
$ I_{6,6}^{(-)}$ cannot be written in terms of the other invariants. $\left({I_{6,6}^{(-)}}\right)^2$ is even under exchange, and can be written in terms of the other invariants.

It is, of course, well-known that $CP$ conservation in the quark sector requires $J=0$, or equivalently, $I_{6,6}^{(-)}=0$. What is new is the structure of the ring of all invariant polynomials, and the relation between the $CP$-conserving and $CP$-violating invariants.

\section{Lepton Invariants for Two Generations}\label{sec:linv2gen}

The structure of the lepton invariants, like the quark invariants, depends on the number of generations, so we first consider the case of $\ng=2$ generations in this section.  The case of $\ng=3$ generations is considered in Section~\ref{sec:linv3gen}. We will outline the derivation of the results, but not give all the details.

\subsection{The Standard Model Effective Theory}\label{sec:sm2}

We now study the lepton invariants in the Standard Model low-energy effective theory with a neutrino Majorana mass term. The structure of the lepton invariants is considerably more complicated than the quark invariants.
The lepton sector of the low-energy theory contains the flavor symmetry breaking matrices $Y_E$ and $C_5$, so we are interested in polynomials in $m_E$, ${m_E}^\dagger$, $m_5$ and ${m_5}^* = {m_5}^\dagger $, since $m_5$ is a symmetric matrix.  These matrices transform as
\begin{eqnarray}
m_E &\to& {\calU_{E^c}}^{T} \ m_E \ \calU_L\ , \nn
{m_E}^\dagger &\to& {\calU_{E^c}}^{\dagger} \ {m_E}^\dagger \ {\calU_L}^*\ , \nn
m_5 &\to& {\calU_{L}}^T \ m_5 \ \calU_L\ ,\nn
{m_5}^* &\to& {\calU_{L}}^\dagger \ {m_5}^* \ {\calU_L}^*\ ,
\end{eqnarray}
under chiral flavor transformations.
To cancel $\calU_{E^c}$, one must consider the combinations
\begin{eqnarray}
X_{E} &\equiv& {m_{E}}^\dagger m_{E} , \nn
X_E^*={X_{E}}^T &\equiv& {m_{E}}^T {m_{E}}^* , 
\end{eqnarray}
which transform as
\begin{eqnarray}
X_{E} &\to& \calU_L^{\dagger} \ X_{E} \ \calU_L,\nn
{X_E}^T &\to& {\calU_L}^T \    {X_E}^T \ {\calU_L}^* \ .
\end{eqnarray}
It also is convenient to define
\begin{eqnarray} 
X_5 \equiv {m_5}^* m_5,
\end{eqnarray} 
which transforms as
\begin{eqnarray}
X_5 &\to& {\calU_L}^\dagger \ X_5 \ \calU_L \ ,
\end{eqnarray}
as well as $\left( {m_5}^* \ \left({X_E}^n\right)^T \ {m_5} \right)$, which transforms as
\begin{eqnarray}
\left( {m_5}^* \ \left({X_E}^n\right)^T \ {m_5}\right) &\to& {\calU_L}^\dagger \left( {m_5}^* \ \left({X_E}^n\right)^T \ {m_5}\right) \ {\calU_L} .\nn
\end{eqnarray}

The invariants involving only $X_E$ are $I_{2,0} = \tr{ X_E }$ and $I_{4,0} = \tr{ {X_E}^2 }$, whereas the invariants involving only $m_5$ and ${m_5}^*$ are $I_{0,2} = \tr{ X_5 }$ and $I_{0,4} = \tr{ {X_5}^2 }$.

The invariants involving $X_E$, $m_5$ and ${m_5}^*$ are of the form
\begin{eqnarray}\label{invleetform2gen}
\tr{  {m_5}^*\  \left( {X_E}^{r_1} \right)^T \ m_5 \  { X_E }^{s_1} \ \ldots  {m_5}^*\  \left( {X_E}^{r_n} \right)^T \ m_5 \  { X_E }^{s_n} }\nn
\end{eqnarray}
for integers $r_i$ and $s_i$.  The Cayley-Hamilton theorem implies that all powers $r_i$ and $s_i$ greater than one in Eq.~(\ref{invleetform2gen}) can be rewritten in terms of lower order invariants.  Thus, one needs to consider traces of matrix products containing the matrices $X_E$, $X_5$, and $\left({m_5}^* \ {X_E}^T \ m_5\right)$ at most once.

In summary, the generators of the invariants are:
\begin{eqnarray}
I_{2,0} &=& \tr{ X_E } = \tr{ {m_E}^\dagger m_E   },  \nn
I_{0,2} &=& \tr{ X_5 } = \tr{ {m_5}^* m_5}, \nn
I_{4,0} &=&\tr{ {X_E}^2} = \tr{ \left({m_E}^\dagger m_E \right)^2 },  \nn
I_{2,2} &=& \tr{ {m_5}^* \  {X_E}^T  \ {m_5} } = \tr{ m_5 \ X_E \ {m_5}^*}\nn
&=& \tr{{m_E}^T \ {m_E}^*  \ {m_5} \ {m_5}^*}
= \tr{ {m_E}^\dagger \ m_E \ {m_5}^* \ {m_5} }, \nn
I_{0,4} &=& \tr{ {X_5}^2 }= \tr{ \left( {m_5}^* m_5 \right)^2 }, \nn
I_{4,2} &=& \tr{  {m_5}^* \ {X_E}^T \ m_5 \ X_E} \nn
&=&  \tr{  {m_5}^* \ {m_E}^T {m_E}^* \ m_5 \ {m_E}^\dagger m_E}, \\
I_{4,4}^{(-)} &=&  \tr{{m_5}^* \ {X_E}^T \ m_5 \ X_E \ {m_5}^* \ m_5 } \nn
&&- \tr{{m_5}^* \ {X_E}^T \ m_5 \ {m_5}^* \ m_5 \ X_E }\nn
&=& \tr{{m_5}^* \ {m_E}^T {m_E}^* \ m_5 \ {m_E}^\dagger m_E  \ {m_5}^* \ m_5 }\nn
&&- \tr{{m_5}^* \ {m_E}^T {m_E}^*  \ m_5 \ {m_5}^* \ m_5 \ {m_E}^\dagger m_E  } , \nonumber
\label{eq100}
\end{eqnarray}
where $I_{4,4}^{(-)}$ is $CP$ odd, and the rest are $CP$ even.  The square of the $CP$-odd invariant, $\left(I_{4,4}^{(-)}\right)^2$, is not independent; it can be expressed in terms of polynomials in the other $CP$-even invariants.  In addition, the $CP$-even invariant $I_{4,4}^{(+)}$, obtained by the substitution $- \to +$ in $I_{4,4}^{(-)}$,  is not independent, and thus is not included in the above list.  

There are six parameters: four masses, one angle and one phase, see Table~V. The four masses, one mixing angle, and one phase, can be determined from $I_{2,0}$, $I_{4,0}$, $I_{0,2}$, $I_{0,4}$, $I_{2,2}$ and $I_{2,4}$ up to a sign ambiguity in the phase, just as for the case of three generations of quarks already discussed. The sign of the phase is fixed by the sign of $I_{4,4}^{(-)}$.

The multi-graded Hilbert series is
\begin{widetext}
\begin{eqnarray}
h(y,z) &=& \frac{1+y^4z^4}{(1-y^2)(1-y^4)(1-z^2)(1-z^4)(1-y^2z^2)(1-y^4z^2)},
\end{eqnarray}
\end{widetext}
where $y$ counts powers of $m_E$ and $z$ counts powers of $m_5$. The single variable Hilbert series is
\begin{eqnarray}
H(q)=h(q,q) &=& \frac{1+q^8}{(1-q^2)^2(1-q^4)^3(1-q^6)}.
\label{hilb3}
\end{eqnarray}
The $q^8$ term in the numerator shows that there is one degree-eight invariant $I_{4,4}^{(-)}$ which occurs, but that the square of this invariant is not independent and can be eliminated.

The number of denominator factors $p=6$ is equal to the number of parameters, and $\dn=8$, $\dd=22$. The number of variables is $\dim V=14$, since we have one $2\times 2$ mass matrix, one $2\times 2$ symmetric mass matrix, and their complex conjugates. Knop's inequality $14 \ge 22-8 \ge 6$ is satisfied, with an equality for the upper bound.  The six parameters correspond to 2 charged lepton masses, 2 Majorana neutrino masses, one mixing angle and one phase.

The denominator of Eq.~(\ref{hilb3}) shows that there are two generators of degree two,  three of degree four, and one of degree six, which agrees with the $CP$-even invariants in  Eq.~(\ref{eq100}). The numerator shows that there is an invariant of degree eight, whose square can be eliminated, which is $I_{4,4}^{(-)}$.
The structure of the invariants for $\ng=2$ is similar to that for quarks for $\ng=3$.

Weak-basis invariants for two generations in the low-energy effective theory were studied previously by Branco, Lavoura and Rebelo~\cite{Branco4}. They defined an invariant $Q$, related to $I_{4,4}^{(-)}$  by
\begin{eqnarray}
2 i\, \text{Im}\, \text{Tr}\, Q &=& I_{4,4}^{(-)}\, ,
\end{eqnarray}
and showed that $Q=0$ is a necessary and sufficient condition for $CP$ conservation. This is consistent with our results, since the only $CP$-odd generating invariant is $I_{4,4}^{(-)}$.

\subsection{The Seesaw Model}\label{sec:all2}

In this section, we analyze the lepton invariants in the seesaw theory for $\ng=\mg=2$ generations of fermions.  There are three matrices in the lepton sector, $m_\nu$, $m_E$ and $M$, and their complex conjugates ${m_\nu}^\dagger$, ${m_E}^\dagger$ and $M^\dagger = M^*$.\footnote{It is worth emphasizing that in our notation $m_\nu$ refers to the Dirac mass matrix $m_\nu = Y_\nu v/\sqrt{2}$, not the Majorana mass matrix $m_5$ of the effective theory.}  From Eq.~(\ref{trans}), we see that only $m_E$ transforms under 
$\calU_{E^c}$, so it must always occur in the combination
\begin{eqnarray}
X_E &=& {m_E}^\dagger m_E,
\end{eqnarray}
which transforms as 
\begin{eqnarray}
X_E &\to& \calU_L^\dagger \ X_E \ \calU_L 
\end{eqnarray}
under the chiral flavor symmetry transformations. The mass matrices $m_\nu$, $m_\nu^\dagger$, $M$ and $M^*$ transform as
\begin{eqnarray}
m_\nu &\to& {\calU_{N^c}}^{T} \ m_\nu \ \calU_L, \nn
m_\nu^\dagger &\to& {\calU_L}^\dagger \ m_\nu^\dagger \ {\calU_{N^c}}^*, \nn
M &\to& {\calU_{N^c}}^T \ M \ {\calU_{N^c}}, \nn
M^* &\to& {\calU_{N^c}}^\dagger \ M^* \ {\calU_{N^c}}^* \ .
\end{eqnarray}
It is useful to define
\begin{eqnarray}
X_\nu &\equiv& m_\nu^\dagger m_\nu,\nn
Z_\nu &=& m_\nu m_\nu^\dagger,\nn
{Z_\nu}^T &=& {Z_\nu}^* = m_\nu^* {m_\nu}^T, 
\end{eqnarray}
which transform as
\begin{eqnarray}
X_\nu &\to& \calU_L^\dagger \ X_\nu \ \calU_L, \nn
{Z_\nu} &\to& {\calU_{N^c}}^T \  {Z_\nu}  \ {\calU_{N^c}}^*,\nn
{Z_\nu}^T &\to& {\calU_{N^c}}^\dagger \  {Z_\nu}^T \ \calU_{N^c},
\end{eqnarray}
as well as
\begin{eqnarray}
X_N &\equiv& M^* M,\nn
Z_N &=& M M^*,\nn
Z_X &=& m_\nu  \, X_E \, {m_\nu}^\dagger
\end{eqnarray}
which transform as
\begin{eqnarray}
X_N &\to& \calU_{N^c}^\dagger \ X_N \ \calU_{N^c}, \nn
{Z_N} &\to& {\calU_{N^c}}^T \  {Z_N}  \ {\calU_{N^c}}^*,\nn
Z_X &\to&{\calU_{N^c}}^{T} Z_X \ {\calU_{N^c}}^* .
\end{eqnarray}
Note that ${Z_N}^T = {Z_N}^* = X_N$.

The invariants involve three mass matrices, $m_E$, $m_\nu$ and $M$. One first can consider the simpler problem of studying invariants which only depend on two out of the three matrices. The first case, invariants involving only $m_E$ and $m_\nu$, consists of invariants formed from traces of  $X_E$ and $X_\nu$ only, with no insertions of $M$ or $M^*$.  These invariants are the same as the invariants in the quark sector with the substitutions $X_U \to X_\nu$ and $X_D \to X_E$.  The second case, invariants involving only $m_\nu$ and $M$, are invariants which do not contain $X_E$.  These have the same structure as invariants constructed in the low-energy theory, with the replacements $m_5 \to M$, $m_E \to m_\nu^T$, i.e.\ $X_E \to Z_\nu^T$.

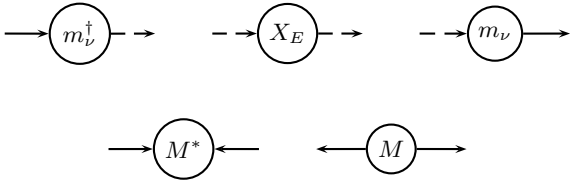
\begin{figure}
\centerline{
\begin{pspicture}(1.5,0.5)(3,1.5)
\rput(2,1){\circlenode{E}{$m_\nu^\dagger$}}
\pnode(1,1){A}
\pnode(3,1){B}
\ncline{->}{A}{E}
\ncline[linestyle=dashed]{->}{E}{B}
\end{pspicture}
\qquad
\begin{pspicture}(1,0.5)(3,1.5)
\rput(2,1){\circlenode{E}{$X_E$}}
\pnode(1,1){A}
\pnode(3,1){B}
\ncline[linestyle=dashed]{->}{A}{E}
\ncline[linestyle=dashed]{->}{E}{B}
\end{pspicture}
\qquad
\begin{pspicture}(1,0.5)(2.5,0.5)
\rput(2,1){\circlenode{E}{$m_\nu$}}
\pnode(1,1){A}
\pnode(3,1){B}
\ncline[linestyle=dashed]{->}{A}{E}
\ncline{->}{E}{B}
\end{pspicture}
}
\vspace{0.5cm}
\centerline{
\begin{pspicture}(1,0.5)(3,1.5)
\rput(2,1){\circlenode{E}{$M^*$}}
\pnode(1,1){A}
\pnode(3,1){B}
\ncline{->}{A}{E}
\ncline{<-}{E}{B}
\end{pspicture}
\qquad
\begin{pspicture}(1,0.5)(3,1.5)
\rput(2,1){\circlenode{E}{$M$}}
\pnode(1,1){A}
\pnode(3,1){B}
\ncline{<-}{A}{E}
\ncline{->}{E}{B}
\end{pspicture}
}
\caption{\label{fig:blobs} Graphical representation of the chiral transformation properties of the lepton mass matrices $X_E$, $m_\nu$, $m_\nu^\dagger$, $M$ and $M^*$.  A solid line represents $\calU_{N^c}$, and a dashed line $\calU_L$. The invariants are obtained by forming graphs with no external lines.}
\end{figure}

The most general invariant involving all three matrices has the structure
\begin{eqnarray}
\tr{M^* A_1 {M} A_2^T \ldots
M^* A_{2n-1}  {M} A_{2n}^T },
\label{eq114}
\end{eqnarray}
where $A_i=\openone$ or $A_i =m_\nu {\cal P}(X_E, X_\nu) {m_\nu}^\dagger$, where ${\cal P}$ is a polynomial in $X_E$ and $X_\nu$.  This result can be obtained by representing the chiral transformations of the matrices graphically, as shown in Fig.~\ref{fig:blobs}.
Products of matrices such as Eq.~(\ref{eq114}) also occurred when studying rephasing invariants~\cite{jmrephasing}. For rephasing invariants, one can factor long products into smaller ones, each involving at most four mixing matrices, using reconnection identities. This 
factorization is no longer possible for the case of mass-matrix invariants, which leads to an interesting and highly non-trivial structure for the invariants.

The basic invariants can be constructed using Eq.~(\ref{eq114}) and eliminating higher powers of matrices by the Cayley-Hamilton identity Eq.~(\ref{a2}).
The generators are:
\begin{eqnarray}
I_{2,0,0} &=& \tr{ X_E} = \tr{ {m_E}^\dagger m_E} ,\nn
I_{0,2,0} &=& \tr{ X_\nu }= \tr{ {m_\nu}^\dagger m_\nu} ,\nn
I_{0,0,2} &=& \tr{X_N}= \tr{ M^* M} ,\nn
I_{4,0,0} &=& \tr{ {X_E}^2 }= \tr{ {m_E}^\dagger m_E {m_E}^\dagger m_E},\nn
I_{2,2,0} &=& \tr{X_\nu  {X_E} }= \tr{ {m_\nu}^\dagger m_\nu {m_E}^\dagger m_E },\nn
I_{0,4,0} &=& \tr{{X_\nu}^2} = \tr{ {m_\nu}^\dagger m_\nu {m_\nu}^\dagger m_\nu} , \nn
I_{0,2,2} &=& \tr{ Z_\nu   Z_N} = \tr{  m_\nu {m_\nu}^\dagger {M} M^* },\nn
I_{0,0,4} &=& \tr{{X_N}^2} = \tr{ M^* M M^* M},\nn
I_{2,2,2} &=& \tr{ Z_X Z_N } = \tr{ m_\nu {m_E}^\dagger m_E {m_\nu}^\dagger M M^* },\nn
I_{0,4,2}&=& \tr{M^* Z_\nu M {Z_\nu}^T }= \tr{  M^* m_\nu {m_\nu}^\dagger M {m_\nu}^* 
{m_\nu}^T  },\nn
I_{2,4,2} &=& \tr{ M^* Z_\nu M {Z_X}^T} \nn 
&=& \tr{  M^* m_\nu {m_\nu}^\dagger M {m_\nu}^*  {m_E}^T {m_E}^* {m_\nu}^T  },\nn
I_{2,4,2}^{(-)} &=& \tr{M^* Z_\nu Z_X M} - \tr{ M^* Z_X Z_\nu M} \nn
&=& \tr{M^* m_\nu {m_\nu}^\dagger m_\nu {m_E}^\dagger m_E {m_\nu}^\dagger M} \nn &&- \tr{ M^* m_\nu {m_E}^\dagger m_E {m_\nu}^\dagger m_\nu {m_\nu}^\dagger M} ,\nn
I_{0,4,4}^{(-)} &=& \tr{ Z_N Z_\nu M  {Z_\nu}^T M^*} - \tr{M^* Z_\nu Z_N M {Z_\nu}^T } \nn
&=& \tr{M M^* m_\nu {m_\nu}^\dagger  M {m_\nu}^* 
{m_\nu}^T  M^*} \nn
&&- \tr{M^* m_\nu {m_\nu}^\dagger  M M^* M {m_\nu}^* {m_\nu}^T  } , \nn
I_{4,4,2}&=&  \tr{M^* Z_X { M} {Z_X}^T }\nn
&=& \tr{  M^* m_\nu {m_E}^\dagger m_E {m_\nu}^\dagger M {m_\nu}^*  {m_E}^T {m_E}^* {m_\nu}^T  },\nn
I_{2,4,4}^{(-)} &=& \tr{Z_N Z_X { M}  {Z_\nu}^T M^*}-  \tr{M^* Z_X Z_N M  {Z_\nu}^T    } \nn
&=& \tr{MM^* m_\nu {m_E}^\dagger m_E {m_\nu}^\dagger { M}  {m_\nu}^* 
{m_\nu}^T M^*   }\nn  &&-  \tr{M^* m_\nu {m_E}^\dagger m_E {m_\nu}^\dagger  M { M}^\dagger M  {m_\nu}^* {m_\nu}^T } ,\nn
I_{2,6,2}^{(-)} &=&  \tr{M^* Z_\nu Z_X {M} {Z_\nu}^T    } - \tr{M^* Z_X Z_\nu { M}  {Z_\nu}^T   } \nn
&=&  \tr{M^* m_\nu {m_\nu}^\dagger m_\nu {m_E}^\dagger m_E {m_\nu}^\dagger {M} {m_\nu}^* 
{m_\nu}^T  } \nn &&- \tr{M^* m_\nu {m_E}^\dagger m_E {m_\nu}^\dagger m_\nu {m_\nu}^\dagger { M}  
{m_\nu}^* {m_\nu}^T  } ,\nn
I_{4,4,4}^{(-)}&=&\vev{M^* Z_N Z_X M Z_X^T} - \vev{M^* Z_X^T Z_N M Z_X},\nn
I_{4,6,2}^{(-)}&=&\vev{M^* Z_\nu Z_X M Z_X^T} - \vev{M^* Z_X Z_\nu M Z_X^T} . \nn
\label{eq115}
\end{eqnarray}
There are several invariants which can be immediately eliminated because they are polynomials in lower order invariants and which have not been listed above.  These invariants  include $I_{2,4,2}^{(+)}$, $I_{0,4,4}^{(+)}$, $I_{2,4,4}^{(+)}$, $I_{2,6,2}^{(+)}$, 
$I_{4,4,4}^{(+)}$ and $I_{4,6,2}^{(+)}$, which are related in an obvious way to the invariants in Eq.~(\ref{eq115}) with superscripts $(-)$.
The degree-eight invariants
$I_{2,4,2}^{(+)}$ and $I_{0,4,4}^{(+)}$ are eliminated by the identities
\begin{eqnarray}
0 &=& I_{0,0,2} I_{0,2,0}^2 I_{2,0,0}-I_{0,0,2} I_{0,4,0} I_{2,0,0}-2I_{0,2,0}I_{2,2,2}\nn
&&-2I_{0,2,2}I_{2,2,0}+2 I_{2,4,2}^{(+)},\nn
0 &=& I_{0,0,2}^2 I_{0,2,0}^2-2 I_{0,0,2} I_{0,4,2}-I_{0,0,4}I_{0,2,0}^2\nn
&&-2I_{0,2,2}^2+2 I_{0,4,4}^{(+)},
\end{eqnarray}
and the degree-ten invariants $I_{2,4,4}^{(+)}$ and $I_{2,6,2}^{(+)}$ are eliminated by the identities
\begin{eqnarray}
0 &=& I_{0,0,2}^2 I_{0,2,0} I_{2,2,0}-2 I_{0,0,2} I_{2,4,2}-I_{0,0,4} I_{0,2,0} I_{2,2,0}\nn
&&-2 I_{0,2,2} I_{2,2,2} +2 I_{2,4,4}^{(+)},\nn
0 &=& I_{0,2,0}^2 I_{0,2,2} I_{2,0,0}-2 I_{0,2,0} I_{2,4,2}-I_{0,2,2} I_{0,4,0} I_{2,0,0}\nn
&&-2 I_{0,4,2} I_{2,2,0}+2 I_{2,6,2}^{(+)}.
\end{eqnarray}
The degree-twelve invariants $I_{4,4,4}^{(+)}$ and $I_{4,6,2}^{(+)}$
are also polynomials in lower order invariants, but we do not include the explicit identities here.

In Eq.~(\ref{eq115}), there are three $CP$-even invariants of degree two, five of degree four, two of degree six, one of degree eight, and one of degree ten, for a grand total of 12 basic $CP$-even invariants.  In addition, there are two $CP$-odd invariants of degree eight, two of degree ten and two of degree twelve, for a total of 6 basic $CP$-odd invariants. All of the invariants can be written as polynomials in these 18 basic invariants.

The multi-graded Hilbert series is 
\begin{widetext}
\begin{eqnarray}
 h(x,y,z)&=&\frac{N}{D},\nn
N&=&1+2 x^2 y^4 z^2 
+ y^4 z^4 +x^2 y^4 z^4 
+x^2 y^6 z^2 
+x^4 y^4 z^4 +x^4 y^6 z^2 
-x^2 y^6 z^6 -x^2 y^8 z^4 
-x^4 y^6 z^6 -x^4 y^8 z^4 \nn
&&-x^6 y^8 z^4 -2 x^4 y^8 z^6 
-x^6 y^{12} z^8 ,\nn
D&=&
   \left(1-x^2\right)\left(1-x^4\right) \left(1-y^2\right)\left(1-y^4\right)
    \left(1-z^2\right) \left(1-z^4\right) \left(1-x^2 y^2\right) 
   \left(1-y^2 z^2\right) \left(1-x^2 y^2 z^2\right)
   \left(1-y^4 z^2\right) \left(1-x^4 y^4 z^2\right),
  \nn
\end{eqnarray}
\end{widetext}
where $x$, $y$, $z$ count powers of $m_E$, $m_\nu$ and $M$, respectively. The Hilbert series $H(q)=h(q,q,q)$ is
\begin{eqnarray}
H(q) &=& \frac{1+q^6+3q^8+2q^{10}+3q^{12}+q^{14}+q^{20}}{(1-q^2)^3(1-q^4)^5(1-q^6)(1-q^{10})}\,,\nn
\label{hilb6}
\end{eqnarray}
which has a palindromic numerator. The number of denominator factors $p=10$ is equal to the number of parameters, and $\dn=20$ and $\dd=42$. The number of variables is $\dim V=22$, because we have two $2 \times 2$ matrices with 4 independent entries, one $2\times 2$ symmetric matrix with 3 independent entries, and their complex conjugates. Knop's inequality is $22 \ge 42-20 \ge 10$, and the upper bound is an equality.  The 10 parameters in the lepton sector of the seesaw model for $\ng=\mg=2$ generations correspond to 2 charged lepton masses, 4 Majorana neutrino masses of the two light and the two heavy neutrinos, 2 angles and 2 phases.

One can see from the Hilbert series that the structure of invariants is far more complicated than in the quark case. The denominator factors $(1-q^2)^3(1-q^4)^5$ of Eq.~(\ref{hilb6}) corresponds to the generators $I_{2,0,0}$, $I_{0,2,0}$,  $I_{0,0,2}$, $I_{4,0,0}$, $I_{2,2,0}$, $I_{0,4,0}$, $I_{0,2,2}$, $I_{0,0,4}$. At degree six, in addition to products of lower order invariants, there are two new invariants, $I_{2,2,2}$ and $I_{0,4,2}$. These two invariants correspond to  the $(1-q^6)$ factor in the denominator, and the $+q^6$ term in the numerator. Since there is only one power of $(1-q^6)$ factor in the denominator, we know that there will be non-trivial relations involving the degree-six invariants. At degree eight, there are 3 new invariants from the $+3 q^8$ term in the numerator in addition to products of lower degree invariants which make up the denominator. These are the three degree-eight invariants in Eq.~(\ref{eq115}). There are three new invariants of degree twelve (from the $+3 q^{12}$), but only two degree-twelve invariants in Eq.~(\ref{eq115}). The third degree-twelve invariant is the square of the degree-six invariant corresponding to the $+q^6$ term in the numerator, so 
the square of this $CP$-even invariant cannot be removed. We have noted earlier that there must be non-trivial relations involving the degree-six invariants. These relations first occur at degree 14,
\begin{eqnarray}
0 &=& I_{0,0,2} I_{0,2,0} I_{2,6,2}^{(-)}+I_{0,2,0}^2 I_{0,4,4}^{(-)} I_{2,0,0}-I_{0,2,0}^2 I_{2,4,4}^{(-)}\nn
&&-I_{0,2,0} I_{0,2,2} I_{2,4,2}^{(-)}-I_{0,2,0} I_{0,4,4}^{(-)} I_{2,2,0}-2 I_{0,2,2} I_{2,6,2}^{(-)}\nn
&&-I_{0,4,0}I_{0,4,4}^{(-)} I_{2,0,0} +2 I_{0,4,0} I_{2,4,4}^{(-)} +2 I_{0,4,2}  I_{2,4,2}^{(-)} \nn
0 &=& I_{0,0,2}^2I_{0,2,0} I_{2,4,2}^{(-)}- I_{0,0,2}^2 I_{2,6,2}^{(-)}+I_{0,0,2} I_{0,2,0} I_{2,4,4}^{(-)}\nn
&&-I_{0,0,2} I_{0,2,2}  I_{2,4,2}^{(-)}-I_{0,0,2}I_{2,2,0} I_{0,4,4}^{(-)}
-I_{0,0,4} I_{0,2,0}  I_{2,4,2}^{(-)}\nn
&&+2 I_{0,0,4}  I_{2,6,2}^{(-)}-2 I_{0,2,2} I_{2,4,4}^{(-)}+2 I_{2,2,2} I_{0,4,4}^{(-)},
\label{eq121}
\end{eqnarray}
and are non-linear relations involving the two degree-six invariants. One can proceed to higher degrees --- there are six relations of degree 16, etc., and verify the number of independent invariants at each degree agrees with Eq.~(\ref{hilb6}). The details of the relations are not important. The main purpose of giving Eq.~(\ref{eq121}) is to show that there can be non-linear relations among the generating invariants. To completely unravel all of the non-linear relations requires going beyond degree 20, the highest power of $q$ in the numerator of Eq.~(\ref{hilb6}).

\section{Lepton Invariants for Three Generations}\label{sec:linv3gen}

In this section, we consider the lepton invariants in the low-energy and high-energy theories for three generations of fermions. The number of invariants is far greater than for two generations, and there are many relations between them. For the low-energy theory, we give the Hilbert series, and the invariants which correspond to the denominator factors. For three generations, even the Hilbert series proved too difficult to compute. For this case, we make some general remarks, and discuss
 some invariants considered previously by Branco et al.~\cite{Branco3,Branco4}, and by Davidson and Kitano~\cite{kitano}.

\subsection{The Standard Model Effective Theory}

The invariants involving only $X_E$ are $I_{2,0} = \tr{ X_E }$, $I_{4,0} = \tr{ {X_E}^2 }$ and $I_{6,0} = \tr{ {X_E}^3 }$, whereas the invariants involving only $m_5$ and ${m_5}^*$ are $I_{0,2} = \tr{ X_5 }$, $I_{0,4} = \tr{ {X_5}^2 }$ and $I_{0,6} = \tr{ {X_5}^3 }$.

The invariants involving $X_E$, $m_5$ and ${m_5}^*$ are of the form
\begin{eqnarray}\label{invleetform}
\tr{  {m_5}^*\  \left( {X_E}^{r_1} \right)^T \ m_5 \  { X_E }^{s_1} \ \ldots  {m_5}^*\  \left( {X_E}^{r_n} \right)^T \ m_5 \  { X_E }^{s_n} }\nn
\end{eqnarray}
for integers $r_i$ and $s_i$.  The Cayley-Hamilton theorem implies that all powers $r_i$ and $s_i$ greater than two in Eq.~(\ref{invleetform}) can be rewritten in terms of lower order invariants.  Thus, one needs to consider traces of matrix products containing the matrices $X_E$, $X_5$, $\left({m_5}^* \ {X_E}^T \ m_5\right)$, and $\left({m_5}^* \ \left({X_E}^2\right) ^T \ m_5\right)$ at most twice. Identity Eq.~(\ref{eq83}) cannot be used to eliminate traces with multiple powers of $m_5$, because $\vev{m_5 A m_5 B}$ gets converted to traces of the form $\vev{m_5^2 A B}$ which are no longer invariant. There are many basic invariants, which involve a single trace, up to degree $m_5^{10} m_E^{12}$, and we do not list them all here. The ones up to degree twelve, which are sufficient for the denominator of the Hilbert series (and hence to determine the parameters) are:
\begin{eqnarray}
I_{2,0} &=& \tr{ X_E } = \tr{ {m_E}^\dagger m_E },  \nn
I_{0,2} &=& \tr{ X_5 } = \tr{ {m_5}^* m_5},  \nn
I_{4,0} &=& \tr{ {X_E}^2} = \tr{ \left( {m_E}^\dagger m_E \right)^2 },  \nn
I_{2,2} &=& \tr{ X_E X_5} = \tr{ {m_E}^\dagger m_E {m_5}^* m_5},\nn
I_{0,4} &=& \tr{ {X_5}^2 }= \tr{ \left( {m_5}^* m_5\right)^2 }, \nn
I_{6,0} &=& \tr{ {X_E}^3}= \tr{ \left( {m_E}^\dagger m_E \right)^3 }, \nn 
I_{4,2}^\prime &=& \tr{ {X_E}^2 X_5}= \tr{ \left({m_E}^\dagger m_E \right)^2  {m_5}^* m_5}, \nn
I_{4,2} &=& \tr{  {m_5}^* \  {X_E}^T \ m_5 \ X_E }\nn
&=&  \tr{  {m_5}^* \   {m_E}^T {m_E}^*  \ m_5 \  {m_E}^\dagger m_E }, \nn
I_{2,4} &=& \tr{ X_E {X_5}^2} = \tr{ {m_E}^\dagger m_E \left({m_5}^* m_5 \right)^2 },\nn
I_{0,6} &=& \tr{ {X_5}^3 }= \tr{ \left( {m_5}^* m_5 \right)^3 }, \nn
I_{6,2} &=& \tr{ {m_5}^* \ {X_E}^T \ m_5 \ {X_E}^2 } \nn
&=& \tr{ {m_5}^* \  {m_E}^T {m_E}^*  \ m_5 \ \left( {m_E}^\dagger m_E \right)^2 } , \nn
I_{4,4}^{(\pm)} &=& \tr{{m_5}^* \ {X_E}^T \ m_5 \ {m_5}^* \ m_5 \ X_E }\nn
&&\pm \tr{{m_5}^* \ m_5 \ {m_5}^* \ {X_E}^T \ m_5 \  X_E} \nn
&=& \tr{{m_5}^* \   {m_E}^T {m_E}^* \ m_5 \ {m_5}^* \ m_5 \  {m_E}^\dagger m_E }\nn
&&\pm \tr{{m_5}^* \ m_5 \ {m_5}^* \   {m_E}^T {m_E}^*  \ m_5 \ {m_E}^\dagger m_E }, \nn
I_{8,2} &=& \tr{ {m_5}^* \ ({X_E}^T)^2 \ m_5 \ {X_E}^2 } \nn
&=& \tr{ {m_5}^* \ \left({m_E}^T {m_E}^* \right)^2 \ m_5 \ \left( {m_E}^\dagger m_E \right)^2 }, \nn
I_{6,4}^{(\pm)} &=& \tr{{m_5}^* \ {X_E}^T  \ m_5 \ {m_5}^* \ m_5 \ {X_E}^2 }\nn &&\pm \tr{{m_5}^* \ m_5 \ {m_5}^* \ {X_E}^T \ m_5 \ {X_E}^2 \ } \nn
&=& \tr{{m_5}^* \  {m_E}^T {m_E}^*   \ m_5 \ {m_5}^* \ m_5 \ \left( {m_E}^\dagger m_E \right)^2 }\nn &&\pm \tr{{m_5}^* \ m_5 \ {m_5}^* \  {m_E}^T {m_E}^*  \ m_5 \ \left( {m_E}^\dagger m_E \right)^2 }, \nn
I_{8,4}^{(\pm)} &=& \tr{{m_5}^* \ ({X_E}^T)^2 \ m_5 \ {m_5}^* \ m_5 \ {X_E}^2 \ } \nn &&\pm \tr{{m_5}^* \ m_5 \ {m_5}^* \ ({X_E}^T)^2 m_5 \ {X_E}^2}\nn
&=& \tr{{m_5}^* \ \left({m_E}^T {m_E}^* \right)^2  \ m_5 \ {m_5}^* \ m_5 \ \left( {m_E}^\dagger m_E \right)^2 } \nn 
&&\pm \tr{{m_5}^* \ m_5 \ {m_5}^* \ \left({m_E}^T {m_E}^* \right)^2 m_5 \ \left( {m_E}^\dagger m_E \right)^2 }.\nn
\label{eq104}
\end{eqnarray}

The multi-graded Hilbert series is
\begin{widetext}
\begin{eqnarray}
h(y,z) &=& \frac{N}{D}, \nn
N&=&-y^{24} z^{18} -2 y^{20} z^{14} -2 y^{20} z^{12} -y^{20} z^{10}
   -2 y^{18} z^{14} -3 y^{18} z^{12} -y^{18} z^{10} -3
   y^{16} z^{14} -3 y^{16} z^{12} -3 y^{16} z^{10} -y^{16} z^8 \nn
   &&-y^{16} z^6 -y^{14} z^{14} -y^{14} z^{12} -y^{14} z^{10} 
   -2 y^{14} z^8 -y^{14} z^6 -y^{12} z^{14} +y^{12} z^4 
   +y^{10} z^{12} +2 y^{10} z^{10} +y^{10} z^8 + y^{10} z^6 \nn
   &&+y^{10} z^4 +y^8 z^{12} +y^8 z^{10} +3 y^8 z^8 +3 y^8 z^6 
   +3 y^8 z^4 +y^6 z^8 +3 y^6 z^6 +2 y^6 z^4 +y^4 z^8 +2 y^4 z^6 
   +2 y^4 z^4 +1,\nn
   D &=& \left(1-y^2\right) \left(1-y^4\right) \left(1-y^6\right) \left(1-z^2\right)
    \left(1-z^4\right) \left(1-z^6\right)\left(1-y^2 z^2\right) \left(1-y^4 z^2\right)^2
 \left(1-y^2 z^4\right)  \left(1-y^6 z^2\right)\nn\nobreak
 &&\times \left(1-y^4
   z^4\right)\left(1-y^8 z^2\right),
\end{eqnarray}
where $y$ counts powers of $m_E$ and $z$ counts powers of $m_5$. The single-variable series $H(q)=h(q,q)$ is
\begin{eqnarray}
H(q)=\frac{1+q^6+2 q^8+4 q^{10}+8 q^{12}+7 q^{14}+9 q^{16}
+10 q^{18}+9 q^{20}+7 q^{22}+8 q^{24}+4 q^{26}
+2 q^{28}+q^{30}+q^{36}  }
   {\left(1-q^2\right)^2 \left(1-q^4\right)^3
   \left(1-q^6\right)^4 \left(1-q^8\right)^2
   \left(1-q^{10}\right)} .
   \label{hilb4}
\end{eqnarray}
\end{widetext}
The number of denominator factors $p=12$ is equal to the number of parameters, and $\dn=36$ and $\dd=66$. The number of variables is $\dim V=30$, because we have one $3 \times 3$ matrix with 9 independent entries, one $3\times 3$ symmetric matrix with 6 independent entries, and their complex conjugates. Knop's inequality is $30 \ge 66-36 \ge 12$, and the upper bound is an equality. Note that the numerator is palindromic.  The 12 parameters consist of 3 charged lepton masses, 3 Majorana light neutrino masses, 3 angles and 3 phases.   

The Hilbert series Eq.~(\ref{hilb4}) has a complicated numerator, which shows that the structure of the invariant ring is highly non-trivial. From the denominator of Eq.~(\ref{hilb4}), we see that there are two generators of degree two,  three of degree four, four of degree six, two of degree eight, and one of degree 10, which can be multiplied freely, with no relations. These account for most of the invariants in Eq.~(\ref{eq104}), but there remains one $CP$-even invariant each of degrees 6, 10, 12, and one $CP$-odd invariant each of degrees 8, 10, 12. These contribute $q^6+q^8+2 q^{10}+2q^{12}$ to the numerator in Eq.~(\ref{hilb4}). The coefficient of $q^8$ in the numerator of Eq.~(\ref{hilb4}) is 2. Where does the other degree-eight invariant not in Eq.~(\ref{eq104}) come from? The degree-six invariant that corresponds to the numerator factor $q^6$ can be multiplied by either of the two degree invariants, $I_{2,0}$ or $I_{0,2}$, to give two additional degree-8 invariants. One of these can be written as a polynomial in lower order invariants;  the other survives. One can continue this analysis to arbitrarily high order --- the entire invariant structure is encoded in a very compact way in the Hilbert series Eq.~(\ref{hilb4}). An explicit example of the construction just discussed is given in Sec.~\ref{sec:all2} for the high-energy theory with $\ng=2$, which provides a simpler example of an invariant ring with non-trivial relations.

For three generations, Branco, Lavoura and Rebelo~\cite{Branco4} defined four invariants:
\begin{eqnarray}
2 i I_1&=& I_{4,4}^{(-)}\nn
2 i I_2 &=& \vev{X_E m_5^* m_5 m_5^* m_5 m_5^* X_E^T m_5}-\text{c.c.}\nn
2 i I_3 &=& \vev{X_E m_5^* m_5 m_5^* m_5 m_5^* X_E^T m_5 m_5^* m_5}
-\text{c.c.}\nn
2 i I_4 &=& \det \left[ m_5 X_E m_5^* + m_5^* X_E^T m_5 \right]-\text{c.c.}
\end{eqnarray}
of degrees $(4,4)$, $(4,6)$, $(4,8)$ and $(6,6)$, and showed that the vanishing of these invariants implies $CP$ conservation. The $CP$-violating invariants of Eq.~(\ref{eq104}) correspond to the denominator factors of the Hilbert series. There are additional $CP$-violating invariants not listed which correspond to terms in the numerator.

\subsection{The Seesaw Model}

The invariants involve three mass matrices, $m_E$, $m_\nu$ and $M$. One first can consider the simpler problem of studying invariants which only depend on two out of the three matrices. The first case, invariants involving only $m_E$ and $m_\nu$, consists of invariants formed from traces of  $X_E$ and $X_\nu$ only, with no insertions of $M$ or $M^*$.  These invariants are in direct analogy to the invariants of the quark sector with the substitutions $X_U \to X_\nu$ and $X_D \to X_E$.  The second case, invariants involving only $m_\nu$ and $M$, are invariants which do not contain $X_E$.  These have the same structure as invariants constructed in the low-energy theory, with the replacements $m_5 \to M$, $m_E \to m_\nu^T$, i.e.\ $X_E \to Z_\nu^T$.

The most general invariant involving all three matrices has the structure
\begin{eqnarray}
\tr{M^* A_1 {M} A_2^T \ldots
M^* A_{2n-1}  {M} A_{2n}^T },
\label{eq1143gen}
\end{eqnarray}
where $A_i=\openone$ or $A_i =m_\nu {\cal P}(X_E, X_\nu) {m_\nu}^\dagger$, where ${\cal P}$ is a polynomial in $X_E$ and $X_\nu$.  The generating invariants are given by using Eq.~(\ref{eq1143gen}). In this case, there are a very large number of generating invariants. They include all  those discussed earlier in the seesaw theory for two generations, as well as many other.

For $\ng= \mg =3$ generations, there are 21 parameters which consist of 9 masses, 6 angles and 6 phases.  The 9 masses are the 3 charged lepton masses, 3 light Majorana neutrino masses and 3 heavy Majorana neutrino masses.  There are 3 angles in the mixing matrix $V$ and 3 angles in the mixing matrix $W$.  There is one $\delta$-type phase in $V$ and in $W$, two Majorana phases $\Psi^\prime$ in $W$, and 2 phases $\bar \Phi$ which are not removeable when $V$ and $W$ are considered together.

We have been unable to construct the multi-graded and one-variable Hilbert series in this case.  However, it is clear that the structure of the invariant relations is extremely complicated. There are a number of constraints on the form of the one-variable Hilbert series.  The denominator must be a product of $p=21$ factors.  The numerator must be palindromic, and $d_N$ and $d_D$ must satisfy the Knop inequality $48 \ge d_D- d_N \ge 21$ since $\dim V=48$.  The number of variables $\dim V=48$ results because there are two $3 \times 3$ matrices $m_E$ and $m_\nu$ with 9 independent entries each, one $3 \times 3$ symmetric matrix $M$ with 6 independent entries, and the complex
conjugates of the three matrices.

Ref.~\cite{Branco3} defined six invariants in the seesaw theory,
\begin{eqnarray}
2i I_1 &=& \vev{Y_\nu Y_\nu^\dagger M^* M M^* (Y_\nu Y_\nu^\dagger)^T M}-\text{c.c.}\nn
2i I_2 &=& \vev{Y_\nu Y_\nu^\dagger M^* M M^* M M^* (Y_\nu Y_\nu^\dagger)^T M}-\text{c.c.}\nn
2i I_3 &=& \vev{Y_\nu Y_\nu^\dagger M^* M M^* M M^* (Y_\nu Y_\nu^\dagger)^T M M^* M}-\text{c.c.}\nn
\end{eqnarray}
which involve $CP$-violating phases which are relevant for leptogenesis, as well as 
\begin{eqnarray}
2i \tilde I_1 &=& \vev{Y_\nu X_E Y_\nu^\dagger M^* M M^* (Y_\nu X_E Y_\nu^\dagger)^T M}-\text{c.c.}\nn
2i \tilde I_2 &=& \vev{Y_\nu X_E Y_\nu^\dagger M^* M M^* M M^* (Y_\nu X_E Y_\nu^\dagger)^T M}-\text{c.c.}\nn
2i \tilde I_3 &=& \vev{Y_\nu X_E Y_\nu^\dagger M^* M M^* M M^* (Y_\nu X_E Y_\nu^\dagger)^T M M^* M}-\text{c.c.}\nn
\end{eqnarray}
which involve the other phases.

Ref.~\cite{kitano} defines an invariant
\begin{eqnarray}
2iI_1 &=& \vev{ \kappa^\dagger \kappa \kappa^\dagger (Y_\nu^T Y_\nu^*)^{-1}
\kappa (Y_\nu^\dagger Y_\nu)^{-1}}
\end{eqnarray}
for leptogenesis, where $\kappa$ is $m_5$ with factors of the Higgs vacuum expectation value removed. This is not a polynomial  in the basic variables of the seesaw model. It can be related to the invariants considered here using the formul\ae\  given below.

Invariants in the seesaw model can be related to those of the low-energy effective theory. The basic relation is Eq.~(\ref{c5}), which relates the neutrino mass matrices in the seesaw model to the Majorana mass matrix $m_5$ in the low-energy effective theory. Clearly, the relations between the invariants cannot be polynomial, since inverse powers of $M$ are involved, but one can write the low-energy invariants in terms of a rational function of the high-energy invariants. The basic
 identities are:
\begin{eqnarray}
\det A \ A^{-1} &=& \vev{A} - A\nn
\det A &=& \frac12 \vev{A}^2-\frac12 \vev{A^2}
\end{eqnarray}
for $2\times2$ matrices, and
\begin{eqnarray}
\det A \ A^{-1} &=& A^2-A \vev{A} -\frac12 \vev{A^2} + \frac12 \vev{A}^2 \nn
\det A &=& \frac13 \vev{A^3}-\frac12 \vev{A^2}\vev{A}+\frac16 \vev{A}^3
\end{eqnarray}
for $3 \times 3$ matrices, which can be combined with
\begin{eqnarray}
C_5 &=& Y_\nu^T M^{-1} Y_\nu =  Y_\nu^T (M^*M)^{-1}M^* Y_\nu
\end{eqnarray}
to obtain the desired relations using $A=M^*M$, and substituting for $C_5$ (i.e.\ $m_5$) in the expressions for the low-energy invariants. The expressions are valid as long as $\det M^* M \not =0$,  i.e.\ as long as the singlet neutrinos are heavy and the transition to a low-energy effective theory is valid.

\section{Conclusions}\label{sec:conclusions}

We have used the mathematics of invariant theory to classify the independent invariants of the Standard Model effective theory and its high-energy seesaw model and to study the non-trivial structure of relations (syzygies) among the invariant generators.  The complete classification of invariants and the Hilbert series have been obtained for the Standard Model effective theory with a dimension-five Majorana neutrino mass operator.  A complete solution also has been obtained for the renormalizable seesaw model with $\ng = \mg =2$ fermion generations.  The lepton sector of the seesaw model involves three different mass matrices, the charged lepton mass matrix, the Dirac Mass matrix of the weakly-interacting doublet neutrinos and the Majorana mass matrix of the gauge-singlet neutrinos.  The invariant structure is very complicated.  In the case of $\ng =\mg=3$ generations of fermions, we have been unable to find the Hilbert series for the invariant generators, and thus the structure of the syzygy relations for three generations remains an open problem.

\acknowledgements

AM would like to thank Professor Nolan Wallach for extensive discussions on invariant theory. The three-family neutrino problem led to a related computation of invariants of interest to mathematicians~\cite{garsia}.

\bibliography{invariants}

\end{document}